\documentclass[aps,prf,groupedaddress,longbibliography]{revtex4-1}

\usepackage{graphicx}
\usepackage{natbib}
\usepackage{graphicx,epsfig}
\usepackage{amsmath}
\usepackage{graphicx,overpic} %7024 4509 8087 2389 8534 4319
\usepackage{amsfonts,amsmath}
\usepackage{subfigure}
\usepackage{epic}
\usepackage{multirow}%\textbf{}
\usepackage{color}

\begin{document}

%\preprint{}

\title{Suppression of Electroconvective and Morphological Instabilities by an Imposed Cross Flow of the Electrolyte}



\author{
Gaojin Li$^1$, Alex Townsend$^2$, Lynden A. Archer$^1$ and Donald L. Koch}
\email[]{dlk15@cornell.edu}

\affiliation{Robert Frederick Smith School of Chemical and Biomolecular Engineering, Cornell University, Ithaca, NY 14853\\
$^2$Department of Mathematics, Cornell University, Ithaca, NY 14853}


\date{\today}

\begin{abstract}
Electroconvection and its coupling with a morphological instability are important in many applications, including electrodialysis, batteries and fuel cells. In this work, we study the effects of a two-dimensional channel flow on the electroconvective and morphological instabilities using two approaches. In the bulk analysis, we consider the instability of the electroneutral bulk region driven by a second kind electroosmosis slip velocity boundary condition and derive the asymptotic solutions for small and large wavenumbers. In the full analysis, we consider the entire region of the liquid electrolyte and use the ultraspherical spectral method to numerically solve the eigenvalue problems. Both studies show that the imposed flow significantly affects the electroconvective instability. The imposed flow generates a shielding effect by deforming the perturbed ion concentration field and hinders the ion transfer from low- to high- concentration regions which causes the instability. It fully suppresses the electroconvective instability at small wavenumbers and reduces the growth rate of the perturbations at large wavenumbers. The direct effect of the flow on the morphological instability is minor, while the suppression of the electroconvective instability may change the wavenumber of the most unstable mode of the coupled instabilities. For the electroconvective instability, the bulk analysis is qualitatively different from the full analysis at high wavenumbers.  For the morphological instability, good agreement is found between the two studies at both small and large wavenumbers.
\end{abstract}

\pacs{}
%\keywords{}

\maketitle

\section{Introduction}
Electroconvection of liquid electrolyte near an ion-selective surface can be found in many applications: electrodialysis ~\cite[]{rubinstein1997electric}, desalination ~\cite[]{kim2010direct}, sample separation and detection ~\cite[]{pennathur2007free, wang2005million}, batteries with liquid electrolytes ~\cite[]{fleury1993coupling, huth1995role}, and fuel cells ~\cite[]{kjeang2009microfluidic}. This phenomenon is caused by an electrohydrodynamic instability mechanism which involves the coupled effects between ion transport, electric field, fluid flow, and ion reactions at the interfaces. Once the applied voltage is above a critical value, the liquid electrolyte near the ion-selective surface starts to flow and generates steady or unsteady vortices. This phenomenon has been observed near permselective membranes and metal electrode surfaces, and it can also be found near the inlets of the nanofluidic channels connecting two micro-chambers of ion solutions. It enhances the mixing of the ions and generates the so-called overlimiting current. In some applications, such as water desalination and bioanalytical sensors, this effect is favorable since it greatly increases the ion removal rate and the mixing efficiency. While in other applications, such as in advanced batteries with metal anodes or the fast charging Li-ion batteries, the electroconvection interacts with the ``Mullins-Sekerka''-type morphological instability ~\cite[]{mullins1964stability} and magnifies the nonuniform dendritic electrodeposition. It reduces rechargeability of the battery and causes safety issues. Understanding the mechanisms of these instabilities is of crucial importance in order to better utilize or suppress them in different applications.

Purely electroconvective instability near a fixed ion-selective surface has been widely studied in the literature using theories ~\cite[]{rubinstein2000electro, rubinstein2001electro, rubinstein2005electroconvective, zaltzman2007electro}, experiments ~\cite[]{rubinstein2008direct, de2015dynamics} and direct numerical simulations ~\cite[]{demekhin2013direct, demekhin2014three, druzgalski2013direct, druzgalski2016statistical, li2019electroconvection}. Near a surface such as a Nafion membrane, the ions simply pass through the membrane without changing the morphology of the surface.
At a small voltage, the current is sustained by a one-dimensional ion transport by migration and diffusion, and the electrolyte remains qusai-electroneutral except it forms a nanometer-sized non-electroneutral electric double layer near to the membrane. With increasing voltage, the strong depletion of the ions near the membrane forms a nonequlibrium double layer, which has a micrometer-sized extended space charge layer outside the original equilibrium double layer. Inside the space charge layer, the perturbation of the electric force and the osmotic pressure generate an electroosmosis slip velocity. The osmotic slip velocity, which is called second-kind to differentiate from the slip velocity caused by the equilibrium double layer, is the fundamental cause of electroconvection. Rubinstein and coworkers have conducted a series of works to investigate this type of instability using two types of analyses, the bulk analysis and the full analysis. In the first approach, the linear stability analysis is performed only in the electroneutral bulk region, and the contribution of the space charge layer is represented by the slip velocity which is determined by the properties of the bulk region ~\cite[]{rubinstein2000electro, rubinstein2001electro}. This simplification facilitates an analytical treatment and predicts the existence of a critical voltage for the onset of the instability. However, it is only qualitatively correct and its prediction of the critical voltage is substantially lower than the one derived in direct numerical simulations ~\cite[]{demekhin2013direct, druzgalski2013direct}. In the full analysis ~\cite[]{zaltzman2007electro}, the entire electrolyte, including the bulk region and the thin layers near the membrane, is considered by solving the anion and cation conservation equations separately. In this analysis, both the one-dimensional base state solution and the eigenvalue problem for the perturbed equation are numerically solved. The prediction  of the critical voltage for the onset of electroconvection based on the full analysis quantitatively agrees with the direct numerical simulations ~\cite[]{demekhin2013direct}. The full analysis also predicts a stable mode at high enough wavenumber in contrast to the bulk analysis which has the ``short-wave catastrophe'' ~\cite[]{zaltzman2007electro}.
The numerical calculation in the full analysis is challenging because the double layer is much thinner than the bulk region (typically 4-7 orders of magnitude). Typical spectral methods easily become ill-conditioned since they require a large number of grid points to resolve the thin layer. In this work, we will use both methods to build a complete understanding on the electroconvective and morphological instabilities.

On a metal electrode surface, the electrodeposition of the metal ions from their salt solutions causes morphological instabilities and generates ramified structures. The phase transformation from liquid into solid is unstable in a transport-limited process ~\cite[]{mullins1964stability}. Depending on the type of salt, the ion concentration and the applied field, the deposition can have different morphologies, such as fractal, dense branching and needle-like ~\cite[]{sawada1986dendritic, grier1986morphology, argoul1988self}. Chazalviel found that the fast ramified growth is directly related to the formation of a space charge layer ~\cite[]{chazalviel1990electrochemical}. In his study, the deposit is modeled as a comb of rectilinear equally spaced needles of infinitely small thickness. In a steady solution without a flow, the advancing speed of the deposits equals the retreating speed of anions in the applied electric field near the electrode surface. In the presence of fluid flow, the net charges at the tips of the deposits induce vortices which bring more ions to the tips and further amplifies the needle growth ~\cite[]{fleury1992theory}.  This positive feedback between electroconvection and the morphological instability is directly observed in experiments ~\cite[]{fleury1993coupling, huth1995role}. The uncontrolled electroconvection near deposition can also lead to different morphologies under similar depositing conditions and generate depositions of network structures ~\cite[]{wang1994formation}. In applications related to batteries and energy storage, many efforts have been devoted in order to suppress the electroconvective and morphological instabilities and achieve stable deposition. Such examples include coating a thin layer of a cross-linked polymers on the electrode surface ~\cite[]{maletzki1992ion, khurana2014suppression, han2016dendrite}, adding high-molecular-weight polymers into the liquid electrolyte ~\cite[]{wei2018stabilizing}, and adding the polymer-grafted colloids. Our previous study shows that the electroconvection can be stabilized by adding polymers to exert extra drag to the liquid electrolyte ~\cite[]{tikekar2018electroconvection}. Other studies show that the electroconvection can be stabilized by increasing the flow resistance by the buoyancy force ~\cite[]{karatay2016coupling} and the boundary confinement ~\cite[]{andersen2017confinement}. Generally speaking, these results are consistent with the analysis for a Newtonian electrolyte which predicts that the critical voltage for the onset of electroconvection increases with increasing fluid viscosity ~\cite[]{rubinstein2000electro}.

Another method of regulating the electroconvection and deposition is to impose a flow in the liquid electrolyte. Coupling between the electroconvection, deposition and an imposed flow can be found in many applications, such as electrodialysis, rotating disk electrodeposition, flow batteries and electrical microfluidic devices. In these flows, the fluid inertia is negligible and the instability is influenced by the interaction between the imposed flow and the electrohydrodynamic instabilities. In a microchannel flow, Kwak et al. showed that the pressure-driven flow reduces the height of the electroconvective vortices as $d\sim V^{2/3}/U_m^{1/3}$, $V$ is the voltage and $U_m$ the maximum velocity of the imposing flow ~\cite{kwak2013shear}. At a high flow rate, the strong mainstream flow confines the fluctuations of electroconvection in the near-wall region. At high voltage, the interaction between the strong electroconvection and the imposed flow leads to large vortices sweeping downstream along with the imposed flow ~\cite[]{kwak2017sheltering}. In this experiment, the flow is nearly two dimensional due to the small thickness in the direction perpendicular to the flow-electric-field plane. Using the direct numerical simulation, Urtenov et al. studied the electroconvection in a two-dimensional electrodialysis cell ~\cite{urtenov2013basic}. Their results show that the current exhibits three regimes with increasing the voltage, (1) the linear regime at small voltage, (2) the smooth plateau regime featured by small stable vortices and partial desalination along the flow, and (3) the oscillating regime of large average current caused by a strong and unsteady electroconvection. The simulation results are consistent with experimental observations using laser interferometric bands visualization ~\cite[]{nikonenko2016competition}. More discussion on flow-through electrodialysis membrane cells in the overlimiting current regime can be found in a recent review paper ~\cite[]{nikonenko2014desalination}. In a three-dimensional channel, numerical simulations show that the pressure-driven flow induces helical vortices originating from the side walls, and the overlimiting current varies nonmonotonically with the channel width due to the changes in the vortex size and spacing ~\cite[]{pham2016helical}. On a rotating disk, the electrodeposition forms spiral structures following the streamlines near the surface ~\cite[]{hill1978polyethylene}.
All these works show that the imposed flow significantly affects the nonlinear dynamics of electroconvection and electrodeposition, but it is still unclear how the imposed flow affects the linear instabilities that lead to the onset of electroconvection and non-planar deposition.

Electrohydrodynamic instabilities can also be observed at the interface between two fluids of different conductivities in a strong electric field ~\cite[]{lin2004instability, chen2005convective, posner2012electric}. In these studies, the applied electric field acts parallel to the flow, and the diffusive mixing layer between the two fluids plays the role of a space charge layer. At higher voltages, the flow exhibits a sequence of transitions from steady to time-periodic and then to aperiodic, chaotic states ~\cite[]{posner2012electric}, similar to the flows near an ion-selective membrane ~\cite[]{urtenov2013basic, nikonenko2016competition}. Stability analysis shows that the onset of the convective instability occurs when the electroviscous velocity is strong enough to compete with diffusion and disturb the mixing layer ~\cite[]{lin2004instability, chen2005convective}. As we will see later, this mechanism is different from the flow near an ion-selective membrane, where the instability can be suppressed by a strong imposed flow. Interactions between the electrohydrodynamic instability and the imposed flow at high Reynolds numbers  were also studied for their applications in electrostatic precipitator for particulate emission reduction and bioaerosol sampling, electrohydrodynamic pumps and mixers, flow control by injecting ions, and heat transfer enhancement ~\cite[]{white1963industrial}. Zhang and coworkers studied the instabilities of a Poiseuille flow of non-conducting dielectric fluid with a unipolar ion injection at a high Reynolds number ~\cite[]{zhang2015modal, zhang2016weakly, li2019absolute}. In their studies, the ions are directly injected near one of the channel walls to create a layer of large charge density. Their results showed that increasing the strength of the electric field changes the modes of the unstable waves ~\cite[]{li2019absolute}. The instability of the flow is complicated due to the coexistence of two types of instability mechanisms,  electrohydrodynamic and inertial mechanisms.

In this work, we consider the interaction between the electrohydrodynamic instability and an imposed flow which is perpendicular to the applied electric field. The fluid inertia is negligible and the only mechanism of instability is electrohydrodynamic caused by the formation of the space charge layer. Compared to previous studies, a distinct feature of this work is that the double layer and space charge layer near the ion-selective surface are extremely thin compared to the channel width. For example, in a typical aqueous electrolyte of concentration $1-10^3\mathrm{mol}/\mathrm{m}^3$, the thickness of the double layer is around 0.1-1nm, the space charge layer is around 1$\mu$m, and the channel width is typically around 1mm. The smallest and largest length scales differ by 6 orders of magnitude. As mentioned before, the bulk analysis uses a slip velocity to replace the thin layers, but its results are only qualitatively correct. In the full analysis, the widely adopted Chebyshev collocation spectral method is numerically unstable because the discretized eigenvalue problem becomes highly ill-conditioned with more than 100 collocation points. To resolve this issue, we use the ultraspherical spectral method ~\cite[]{olver2013fast} to solve both the base and perturbed equations in the full analysis.

We study the effects of an imposed flow on the electroconvective and morphological instabilities. In most experiments, the instability mainly occurs in the flow-electric field plane because the confinement in the third direction restricts the out-of-plane wave vectors from occurring. Therefore, we only consider the two-dimensional flow instabilities in the flow-electric field plane. To simplify the problem, we neglect the entrance region and only consider the modal instability of the fully developed flow. This simplification is valid when the ion concentration polarization along the flow occurs rapidly under a relative strong electric field. Two problems will be considered, the purely electroconvective instability and the coupled electroconvective and morphological instability, depending on whether the ion-selective surface is fixed or not.   We describe the problem setup and the governing equations in section \ref{sec:problem}.  In section \ref{sec:bulk}, we conduct the bulk analysis and use the resulting analytical solution to gain a physical understanding of the instability. Then, we will consider the full problem by numerically solving the eigenvalue problems in section \ref{sec:full}. A detailed comparison between the two analyses will be performed to evaluate the viability of the bulk analysis. A summary and conclusion are presented in section \ref{sec:conclusion}.

\section{Problem setup and governing equations}\label{sec:problem}

\begin{figure}
\begin{center}
\includegraphics[angle=0,scale=0.5]{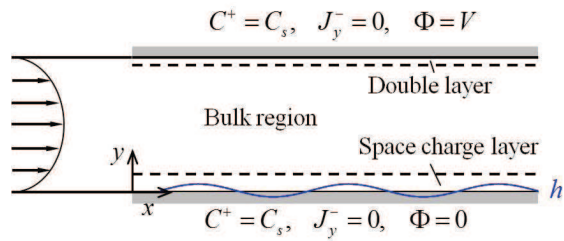}
\caption{Schematic of a  fully developed cross-flow of an electrolyte between two ion-selective surfaces.}\label{fig:fig1}
\end{center}
\end{figure}

\begin{figure}
\begin{center}
\includegraphics[angle=0,scale=0.34]{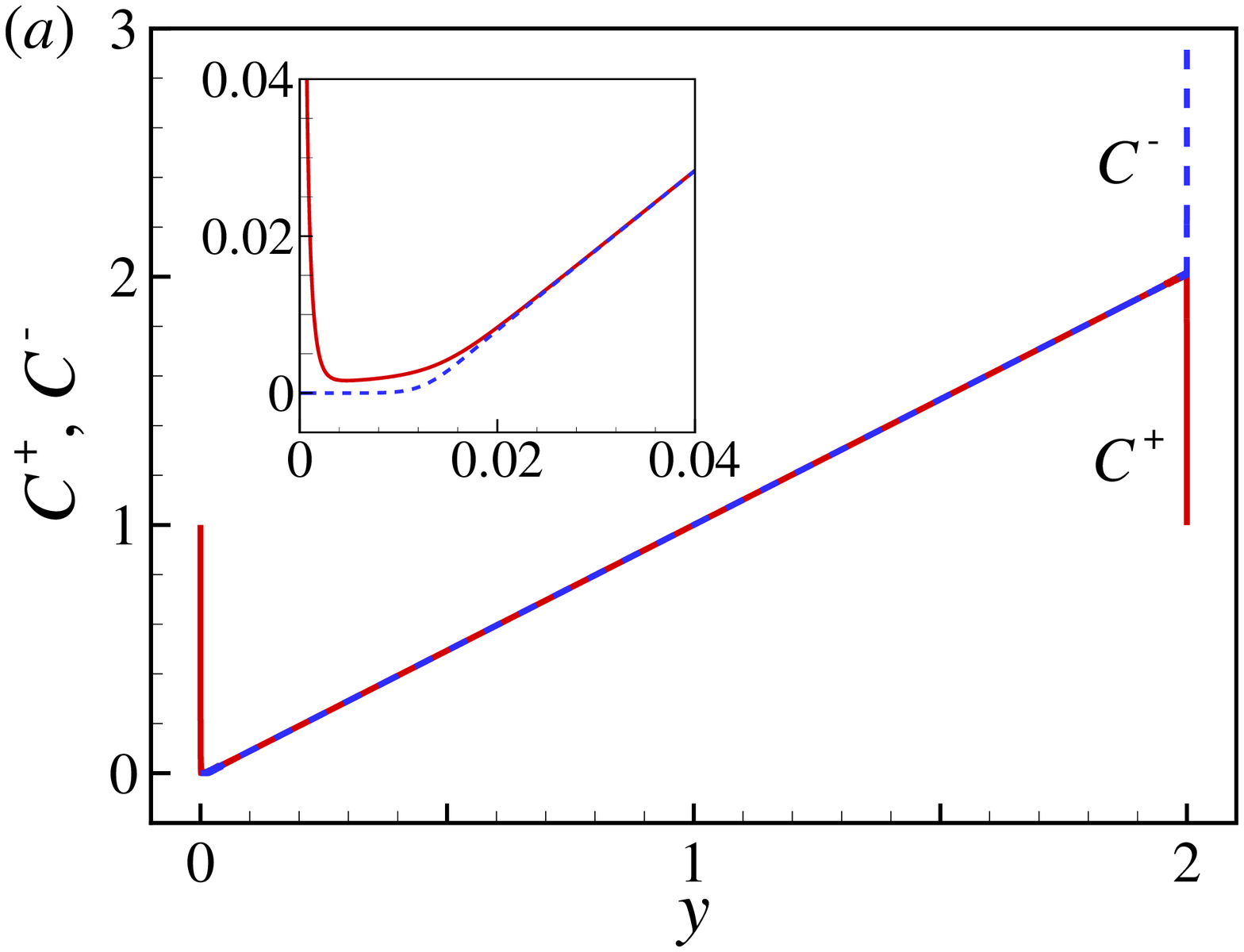}
\includegraphics[angle=0,scale=0.34]{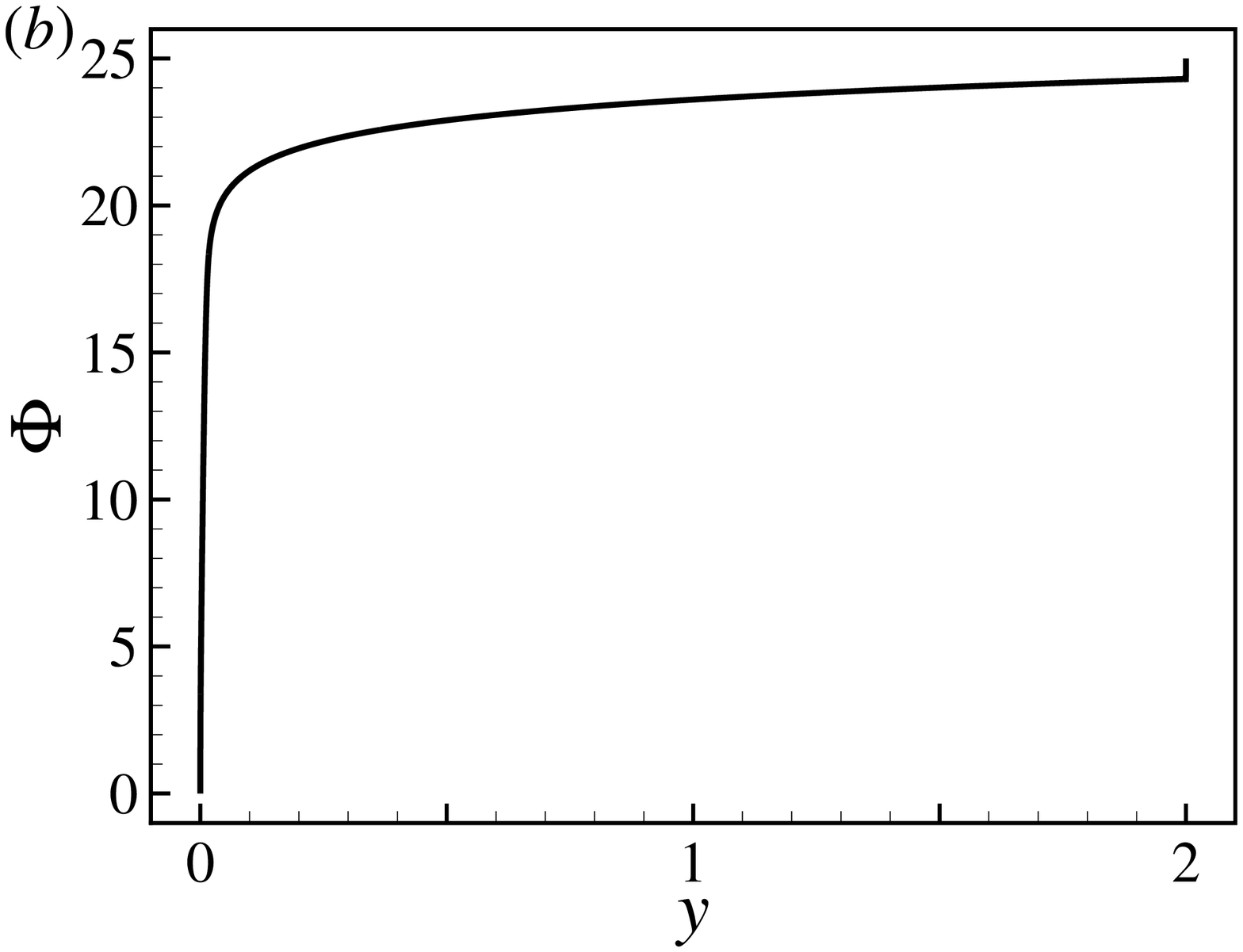}
\caption{Distributions of ($a$) ion concentration and ($b$) potential of the full base state solution at $V=25$, which has a non-equilibrium double layer with a space charge layer at $y=0$, an equilibrium double layer at $y=2$ and a quasi-electroneutral bulk region in between. The bulk analysis only considers the bulk region. The normalized double layer thickness is $\delta=10^{-4}$ and the current is $I=2.0241$.}\label{fig:fig2}
\end{center}
\end{figure}

As shown in figure \ref{fig:fig1}, we consider a pressure-driven flow of a binary univalent electrolyte in a channel of width $2L$. The applied voltage $V$ is perpendicular to the flow direction and buoyancy effects are neglected. The channel walls only allow a non-zero cation flux, and we will consider two types of surfaces,  ion-selective membranes which allow only an electroconvective instability, and metal electrode surfaces with  coupled electroconvective and morphological instabilities. We only consider the morphological instability of the anode surface where ions are deposited, since experiments show that the stripping process on the cathode surface is typically uniform.
Figure \ref{fig:fig2} shows the distribution of ion concentration and potential for fully developed ion transport.
In the base state, the electrolyte inside the channel has three regions at the limiting current: the quasi-electroneutral bulk region with a linear ion concentration profile, the equilibrium double layer at the top surface, and the nonequilibrium double layer which has an extended space charge layer at the bottom surface. We neglect the entrance region of the flow and consider the temporal instability of the fully developed region with a uniform space charge layer.

The governing equations of the problem include the Nernst-Planck equations for the conservation of ion concentration, the Poisson equation for the electrical potential, and the Stokes equation for the incompressible fluid. In non-dimensional form, the equations are
\begin{subequations}\label{eq:full}
\begin{equation}\label{eq:full1}
\frac{\partial C^+}{\partial t}
+Pe(\mbox{\boldmath$u$}\cdot\nabla)C^+
=\frac{1+D}{2}\nabla\cdot(\nabla C^++C^+\nabla\Phi),
\end{equation}
\begin{equation}\label{eq:full2}
\frac{\partial C^-}{\partial t}
+Pe(\mbox{\boldmath$u$}\cdot\nabla)C^-
=\frac{1+D}{2D}\nabla\cdot(\nabla C^--C^-\nabla\Phi),
\end{equation}
\begin{equation}
-2\delta^2\nabla^2\Phi=C^+-C^-,
\end{equation}
\begin{equation}
-\nabla p+\nabla^2\mbox{\boldmath$u$}
+(\nabla^2\Phi)\nabla\Phi=0, \quad \nabla\cdot\mbox{\boldmath$u$}=0,
\end{equation}
\end{subequations}
where $C^+$ and $C^-$ are cation and anion concentrations, $\Phi$ is potential, $p$ pressure and $\mbox{\boldmath$u$}$ fluid velocity. $D=D^+/D^-$ is the ratio of the cation and anion diffusivities, $\delta=\sqrt{\varepsilon\varepsilon_0RT/2F^2C_0}/L$ is the dimensionless double layer thickness, $L$ is the half interelectrode distance, $\varepsilon$ the dielectric constant, $\varepsilon_0$ the vacuum permittivity, $R$ the ideal gas constant, $T$ temperature, $F$ Faraday's constant, and $C_0$ the average ion concentration. $Pe=U_0L/D_0$ is the Peclet number, $U_0=\varepsilon\varepsilon_0(RT/F)^2/\eta L$ is a characteristic velocity derived by
balancing the characteristic Maxwell stress $\varepsilon\varepsilon_0(RT/F)^2$ and the viscous stress $\eta U_0/L$, $\eta$ is the fluid viscosity, $D_0=2D^+D^-/(D^++D^-)$ the average ion diffusivity. Following \cite{rubinstein2005electroconvective}, lengths are non-dimensionalized by $L$, velocity by $U_0$, time by $L^2/D_0$, ion concentration by $C_0$, potential by $RT/F$, and stress by $\eta U_0/L$.

We now discuss the boundary conditions at the two surfaces. In a stripping/plating process between metal electrodes, the cation is stripped from the cathode and then deposited to the anode, therefore the average heights of the two surfaces are continuously changing due to the lithium ion transfer. We choose a reference frame moving with the same velocity as the cathode surface which has uniform stripping. In this frame, the deformation of the anode surface is perturbed around $y=0$, and the cathode surface is fixed at $y=2$. The growth of the anode surface is represented by $y=h(x,t)$. At the two surfaces, the conditions for ion concentration, potential and fluid velocity are
\begin{subequations}\label{eq:fbc}
\begin{equation}\label{eq:fbc1}
C^+|_{y=h}=C_s, \quad
C^+|_{y=2}=C_s', \quad
(\frac{\partial C^-}{\partial y}
-C^-\frac{\partial\Phi}{\partial y})\bigg|_{y=h, 2}
=0,
\end{equation}
\begin{equation}\label{eq:fbc2}
\Phi|_{y=h}=0, \quad
\Phi|_{y=2}=V,
\end{equation}
\begin{equation}\label{eq:fbc3}
\mbox{\boldmath$n$}\cdot\mbox{\boldmath$u$}|_{y=h}=\frac{1}{\mathrm{Pe}}\frac{\partial h}{\partial t}
\frac{1}{|\mbox{\boldmath$n$}\cdot\mbox{\boldmath$e$}_y|}, \quad
\mbox{\boldmath$n$}\cdot\mbox{\boldmath$u$}|_{y=2}=0, \quad
(\textbf{I}-\mbox{\boldmath$n$}\mbox{\boldmath$n$})\cdot\mbox{\boldmath$u$}|_{y=h, 2}=0,
\end{equation}
\begin{equation}\label{eq:fbc4}
\frac{1+D}{2}
(\frac{\partial C^+}{\partial y}+C^+\frac{\partial\Phi}{\partial y})\bigg|_{y=h}-I
=\frac{1}{v_m}\frac{\partial h}{\partial t}\frac{1}{|\mbox{\boldmath$n$}\cdot\mbox{\boldmath$e$}_y|}.
\end{equation}
\end{subequations}
In equ. (\ref{eq:fbc1}), we assume the electrochemical potential of the cation is at equilibrium on the electrode surfaces and therefore the cation concentration is fixed. Its specific value does not qualitatively change the results and $C_s=C_s'=1$ is used in this study. The surfaces are impermeable to anions. Equ. (\ref{eq:fbc2}) is the condition for the applied potential. In equ. (\ref{eq:fbc3}), $\mbox{\boldmath$n$}$ is the unit normal vector of the anode surface pointing into the electrolyte, $\mbox{\boldmath$e$}_y$ is the unit $y$-vector, and $\textbf{I}$ is the identity matrix. The first equation indicates that the growth of the anode generates a normal velocity of the electrolyte, while the other equations are the usual no-slip and no-penetration conditions of the fluid. Equ. (\ref{eq:fbc4}) indicates that the growth rate of the perturbed anode surface is determined by the local cation flux minus the constant base state current $I$.  Here, $v_m=v^*_mC_0$ is the dimensionless molar volume of the metal and $v^*_m$ is the dimensional molar volume. The volume change of the solution during charge/discharge is neglected.
For ion-selective membranes without a morphological instability, $h\equiv0$ and equ. (\ref{eq:fbc4}) becomes a trivial condition.

The governing equations (\ref{eq:full}) with boundary conditions (\ref{eq:fbc}) are first solved for the base state. In the base state, $h=0$, and the pressure-driven flow, ion concentrations and electric field change only along the $y$-direction, so that the equations become
\begin{subequations}\label{eq:fullbase}
\begin{equation}\label{eq:fullbase1}
\frac{1+D}{2}\left(\frac{\partial C^+}{\partial y}
+C^+\frac{\partial\Phi}{\partial y}\right)=I, \quad
\frac{\partial C^-}{\partial y}
-C^-\frac{\partial\Phi}{\partial y}=0,
\end{equation}
\begin{equation}
-2\delta^2\frac{\partial^2\Phi}{\partial y^2}=C^+-C^-,
\end{equation}
\end{subequations}
with boundary conditions
\begin{subequations}\label{eq:fbcbase}
\begin{equation}
C^+|_{y=0}=C_s, \quad
\int_0^2C^-dy=2,
\end{equation}
\begin{equation}
\Phi|_{y=0}=0, \quad
\Phi|_{y=2}=V.
\end{equation}
\end{subequations}
The fluid exhibits Poiseuille flow with $u=U_my(2-y)$ and $v=0$, where $U_m$ is the maximum fluid velocity. Equ. (\ref{eq:fullbase1}) states that in the base state the current $I$ is driven by the cation flux and the anion flux vanishes at any cross-section plane in the electrolyte. In equ. (\ref{eq:fbcbase}), the condition for the anion means that its total concentration is conserved inside the channel due the no flux conditions on both surfaces. The full base state governed by equ. (\ref{eq:fullbase}) requires a numerical solution. Figure \ref{fig:fig2} shows the full base state solutions for ion concentration and potential at steady state at the limiting current. Here, $\delta=10^{-4}$ and $V=25$. The full base state solution has three parts: a non-equilibrium double layer which includes a space charge layer at $y=0$, an equilibrium double layer at $y=2$, and a quasi-electroneutral bulk region. The base state solution of the bulk region will be discussed in section \ref{sec:bulk}.

The bulk analysis deals with the approximate solution of equ. (\ref{eq:fullbase}) by taking advantage of the condition $\delta\ll1$. In a typical electrochemical system, the normalized double layer thickness is $10^{-7}<\delta<10^{-3}$, and the difference between cation and anion concentrations is of order $\delta^2$ in the bulk region. The space charge layer thickness $\delta_s\sim(\delta V)^{2/3}$ is also small as long as the overpotential $V\ll|\ln\delta|$ ~\cite[]{rubinstein2005electroconvective}. In the bulk analysis, the effects of these thin layers are replaced by electroosmotic slip velocities ~\cite[]{rubinstein2000electro, rubinstein2001electro, rubinstein2005electroconvective}. To establish a physical understanding of the effects of the imposed cross flow on the linear instability, we will first discuss the bulk analysis in section \ref{sec:bulk}. The full analysis will later be considered in section \ref{sec:full}.

In typical experiments with aqueous electrolytes, the half interelectrode distance is around $1\mathrm{mm}$, the double layer thickness ranges from $0.1-1\mathrm{nm}$, the dynamic viscosity $\eta=10^{-3}\mathrm{Pa}\cdot\mathrm{s}$, the dielectric constant of water $\varepsilon=80$, the lithium ion diffusivity $10^{-9}\mathrm{m}^2/\mathrm{s}$, ion concentration  $C_0=0.01-1\mathrm{M}$, the lithium ion transference number $t_c=D^+/(D^++D^-)=0.4$, the molar volume of the lithium metal atom $v^*_m=1.3\times10^{-5}\mathrm{m}^3/\mathrm{mol}$, the applied voltage 1-5 Volt, and the typical velocity in a microchannel is up to $10^4\mu \mathrm{m}/\mathrm{s}$. Based on these parameters, we choose $\mathrm{Pe}=0.35, D=0.67, v_m=0.013, U_m=0\sim10^4$ and $\delta=10^{-5}\sim10^{-3}$ for this study.

\section{Bulk analysis}\label{sec:bulk}
In the bulk region, the cation and anion are assumed to have equal concentrations in both the base and perturbed state. Following a previous analysis \cite{rubinstein2005electroconvective}, we introduce the electrochemical potential for the anion $\mu=\ln C-\Phi$ with $C=C^+=C^-$. This treatment avoids directly solving the electrical potential field, which becomes singular at $y=0$ at the limiting current. The governing equations (\ref{eq:full}) become
\begin{subequations}\label{eq:neutral}
\begin{equation}
\frac{\partial C}{\partial t}
+Pe(\mbox{\boldmath$u$}\cdot\nabla)C
=\nabla^2C,
\end{equation}
\begin{equation}
\frac{\partial C}{\partial t}
+Pe(\mbox{\boldmath$u$}\cdot\nabla)C
=\frac{D+1}{2D}\nabla\cdot(C\nabla\mu),
\end{equation}
\begin{equation}
-\nabla p+\nabla^2\mbox{\boldmath$u$}=0, \quad
\nabla\cdot\mbox{\boldmath$u$}=0.
\end{equation}
\end{subequations}
The first two equations are linear combinations of the Nernst-Planck equations (\ref{eq:full1}) and (\ref{eq:full2}) for the cation and anion, the third condition is the Stokes equations without the electric force. The boundary conditions become ~\cite[]{rubinstein2005electroconvective}
\begin{subequations}\label{eq:neutralbc}
\begin{equation}\label{eq:bc1}
C|_{y=h}=0, \quad
(2\ln C-\mu)|_{y=2}=V+\ln C_s,
\end{equation}
\begin{equation}\label{eq:bc2}
\frac{\partial\mu}{\partial y}\bigg|_{y=h, 2}=0,
\end{equation}
\begin{equation}\label{eq:bc3}
v|_{y=h}=\frac{1}{Pe}\frac{\partial h}{\partial t}, \quad
v|_{y=2}=0,
\end{equation}
\begin{equation}\label{eq:bc4}
u|_{y=h}=-\frac{V^2}{8}\frac{\frac{\partial^2C}{\partial y\partial x}}
{\frac{\partial C}{\partial y}}\bigg|_{y=h}, \quad
u|_{y=2}=2\ln2\frac{\partial\mu}{\partial x}\bigg|_{y=2},
\end{equation}
\begin{equation}\label{eq:bc5}
(D+1)\frac{\partial C}{\partial y}\bigg|_{y=h}-I
=\frac{1}{v_m}\frac{\partial h}{\partial t}.
\end{equation}
\end{subequations}
In equ. (\ref{eq:bc1}), the first condition indicates that the anion is completely depleted from the anode surface, the second condition represents the continuity of the chemical potential of the cation across the double layer near the cathode. Equ. (\ref{eq:bc2}) represents the zero-flux condition for anion on both electrodes. Equ. (\ref{eq:bc3}) is the same as its counterpart in (\ref{eq:fbc3}), the growth of the perturbed anode surface equals the normal velocity of the electrolyte, while the cathode surface is fixed in the moving frame of reference. In equ. (\ref{eq:bc4}), the first equation represents the second-kind osmotic slip velocity developed at the edge of the space charge layer near the anode, and the second equation represents the first-kind osmotic slip velocity at the edge of the equilibrium double layer near the cathode. Detailed derivations for the slip velocities are given in previous works \cite[]{rubinstein2000electro, rubinstein2001electro}. Equ. (\ref{eq:bc5}), which is simplified from (\ref{eq:fbc4}) using the condition of equal migration and diffusion fluxes, indicates that the growth rate of the perturbed surface is caused by the perturbation in the cation flux.

As mentioned before, two types of problems will be considered in this section. The first problem is the purely electroconvective instability with $h\equiv0$ and the second problem is a coupled electroconvective and morphological instability. The base state solutions for the two problems are exactly the same. In the rest of the paper, we use capital letters ($C^\pm, M$ and $\Phi$) to represent the base solution, and small letters for the perturbed variables ($c^\pm, \mu, \phi, u, v$ and $h$). By directly solving equ. (\ref{eq:neutral}), it is easy to find the base state of the ion concentration and chemical potential for anion is
\begin{equation}\label{eq:base}
C=y, \quad
M=2\ln2-V,
\end{equation}
with limiting current $I=2$.

We next perform a linear stability analysis to the base state solution. The ion concentration, chemical potential and velocity are perturbed as $c=C+\epsilon  c(y)e^{ikx+\sigma t}, \mu=M+\epsilon\mu(y)e^{ikx+\sigma t}, u=U_my(2-y)+\epsilon u(y)e^{ikx+\sigma t}, v=\epsilon v(y)e^{ikx+\sigma t}$, and the anode surface $h=\epsilon h(y)e^{ikx+\sigma t}$, where $\epsilon\ll1$ is a small perturbation. The perturbed governing equations are
\begin{subequations}\label{eq:perturbed}
\begin{equation}\label{eq:perturbed1}
\sigma c+ikPeU_my(2-y)c+Pev=c''-k^2c,
\end{equation}
\begin{equation}\label{eq:perturbed2}
\sigma c+ikPeU_my(2-y)c+Pev=\frac{D+1}{2D}(y\mu''+\mu'-k^2y\mu),
\end{equation}
\begin{equation}\label{eq:perturbed3}
v^{(4)}-2k^2v''+k^4v=0,
\end{equation}
\end{subequations}
where the prime denotes the derivative with respect to $y$. $k$ is the wavenumber, $\sigma$ is the complex eigenvalue whose real part $\sigma_r$ determines the growth rate of the perturbation and imaginary part determines the wave speed $u_c=-\sigma_i/k$. The pressure-driven flow causes an ion advection along the flow direction. The perturbed boundary conditions on the two surfaces are
\begin{subequations}\label{eq:perturbedbc1}
\begin{equation}\label{eq:perturbedbc1a}
c(0)+h=0, \quad
c(2)-\mu(2)=0,
\end{equation}
\begin{equation}\label{eq:perturbedbc1b}
\mu'(0)=0, \quad
\mu'(2)=0,
\end{equation}
\begin{equation}\label{eq:perturbedbc1c}
v(0)=\frac{\sigma}{Pe}h, \quad
v(2)=0,
\end{equation}
\begin{equation}\label{eq:perturbedbc1d}
v'(0)=-\frac{V^2}{8}k^2c'(0), \quad
v'(2)=2\ln2k^2\mu(2).
\end{equation}
\begin{equation}\label{eq:perturbedbc1e}
c'(0)=\frac{\sigma}{(D+1)v_m}h.
\end{equation}
\end{subequations}
In the following, we will first discuss the purely electroconvective instability and then the coupled electroconvective and morphological instability. For small/large ($k\ll1$ or $k\gg1$) wavenumbers or when the perturbed flow is negligible ($v\ll1$), we will analytically solve the equations (\ref{eq:perturbed}). For arbitrary wavenumber $k$, the equations will be numerically solved using the Chebyshev collocation method ~\cite[]{weideman2000matlab}.

\subsection{Purely electroconvective instability}
In this subsection, we consider the purely electroconvective instability. We will first discuss the asymptotic solutions for the mode with the largest growth rate for $k\ll1$, $k\gg1$ and $v\ll1$. Then, we show the results for arbitrary wavenumbers by numerically solving equations (\ref{eq:perturbed}).

\subsubsection{Small wavenumber, $k\ll1$}
In the limit of small wavenumber ($k\ll1$), the ion concentration and chemical potential are expanded as $c=c_0+kc_1+k^2c_2+\cdots$, $\mu=\mu_0+k\mu_1+k^2\mu_2+\cdots$, the normal velocity $v=k^2v_2+\cdots$ because of the slip velocity condition (\ref{eq:perturbedbc1d}), and the growth rate $\sigma=\sigma_0+k\sigma_1+k^2\sigma_2+\cdots$. The leading order solutions are unaffected by the pressure driven flow, and the results are given by Rubinstein \emph{et al}. ~\cite{rubinstein2005electroconvective}
\begin{equation}
c_0=y, \quad
\mu_0=2, \quad
v_2=\left(\frac{V^2}{16}+(\ln2-\frac{V^2}{32})y\right)(y-2)y,
\end{equation}
with the leading order growth rate $\sigma_0=0$. From the continuity equation $iku+v'=0$, the slip velocity $u_s=u(0)=iV^2k/8$ is of order $k$. This is because the perturbation of the ion concentration occurs over the entire bulk region and it has an O(1) normal gradient.

The effects of the pressure-driven flow arise in higher order equations. For each order, the basic steps are to first calculate $c$ in terms of $\mu$, and then substitute it into equ. (\ref{eq:perturbed2}) and integrate the equation with boundary conditions (\ref{eq:perturbedbc1b}) to derive $\sigma$. In the end, the growth rate is found to be
\begin{equation}\label{eq:smallk1}
\sigma=-\frac{2i}{3}\mathrm{Pe}U_mk
+\left[\mathrm{Pe}(\frac{V^2+32\ln2}{48})-\frac{D+1}{D}
-\frac{8}{945}\mathrm{Pe}^2U_m^2\right]k^2+\cdots.
\end{equation}
This result recovers the previous result ~\cite{rubinstein2005electroconvective} when $U_m=0$. In the presence of the Poiseuille flow, the small wavenumber perturbation propagates with the average fluid velocity $2\mathrm{Pe}U_m/3$, in which the prefactor $\mathrm{Pe}$ exists because time is scaled by $L^2/D_0$ instead of $L/U_0$. The growth rate is reduced at order $k^2$. The O($k$) ion concentration and chemical potential are
\begin{equation}
c_1=i\mathrm{Pe}U_m\left(\frac{D-4}{45(D+1)}y
-\frac{y^3}{9}+\frac{y^4}{6}-\frac{y^5}{20}\right),
\end{equation}
\begin{equation}
\mu_1=-\frac{iD}{D+1}PeU_m(\frac{y^4}{8}-\frac{4y^3}{9}+\frac{y^2}{3}).
\end{equation}
The asymptotic solution for $k\ll1$ arises due to the finite  distance between the two electrode surfaces. It can be either stable or unstable, depending on $Pe, D$ and $U_m$. The imposed flow always reduces the growth rate of the mode for $k\ll 1$. This result is consistent with the full analysis.

\subsubsection{Negligible electroconvection, $v\approx0$}
Because of the pressure-driven flow, the electroconvective velocity is negligible below a critical wavenumber. The eigenmode is solved by $c''=(-a+a^2(y-1)^2)c$ with $a^2=-ikPeU_m$ and $\sigma+k^2=a^2-a$. The growth rate is
\begin{equation}\label{eq:vequ0}
\sigma=-(k^2+\sqrt{k\mathrm{Pe}U_m/2})
+i(-k\mathrm{Pe}U_m+\sqrt{k\mathrm{Pe}U_m/2}),
\end{equation}
i.e., this mode is always stable and its wave speed is the maximum fluid velocity $\mathrm{Pe}U_m$. The ion concentration is
\begin{equation}\label{eq:vequ0c}
c=e^{-\sqrt{\mathrm{Pe}kU_m/8}(y-1)^2(1-i)},
\end{equation}
which is a Gaussian distribution at the centerline of the channel. As we will see in the following, the full analysis has the exact solution for this mode.
This center mode is always stable and it occurs below a transition wavenumber when the imposed flow is strong enough to overcome the electroosmotic slip velocity. Above the transition wavenumber, the electroosmotic slip velocity dominates the imposed flow and the electroconvective instability is determined by the wall modes.

\subsubsection{Large wavenumber, $k\gg1$}
For $k\gg1$, the analysis is performed in the vicinity of the anode surface and the interelectrode distance can be considered as semi-infinite. Introducing the inner length scale $z=ky$ and considering the perturbations which decay away from the anode, i.e., $c(\infty)=v(\infty)=v'(\infty)=0$, the perturbations are expanded as $c=c_0+c_1/k+c_2/k^2+\cdots$ and $\sigma=k^2\sigma_{-2}+k\sigma_{-1}+\sigma_0+\cdots$. In the inner scale, the momentum equation (\ref{eq:perturbed3}) becomes $v^{(4)}-2v''+v=0$, where the prime now denotes the derivative with respect of $z$. Its solution is
\begin{equation}\label{eq:largek1v}
v=k^2ze^{-z}.
\end{equation}
From the continuity equation $iu+v'=0$, the tangential velocity $u=iv'$ and the slip velocity
\begin{equation}\label{eq:slip}
u_s=u(0)=ik^2.
\end{equation}
At high wavenumber, the slip velocity is of order $k^2$ because the ion concentration disturbance is concentrated near the anode surface and has an O($k$) normal gradient.

To calculate the growth rate, we need to solve the equations for the ion concentration. We find that $\sigma_{-2}$ is not influenced by the imposed flow,  $\sigma_{-1}=c_1=0$, $\sigma_0$ and $\sigma_1$ are purely imaginary and one has to solve the fourth order equation for $\sigma_2$ to see the effects of flow on the growth rate. The governing equations and the boundary conditions at each order are
\begin{subequations}
\begin{equation}
c_0''-(1+\sigma_{-2})c_0=\mathrm{Pe}v_{-2},
\end{equation}
\begin{equation}
c_1''-(1+\sigma_{-2})c_1-\sigma_{-1}c_0=0,
\end{equation}
\begin{equation}
c_2''-(1+\sigma_{-2})c_2-\sigma_{-1}c_1-\sigma_0c_0=2i\mathrm{Pe}U_mzc_0,
\end{equation}
\begin{equation}
c_3''-(1+\sigma_{-2})c_3-\sigma_{-1}c_2-\sigma_0c_1-\sigma_1c_0
=i\mathrm{Pe}U_m(2zc_1-z^2c_0),
\end{equation}
\begin{equation}
c_4''-(1+\sigma_{-2})c_4-\sigma_{-1}c_3-\sigma_0c_2-\sigma_1c_1-\sigma_2c_0
=i\mathrm{Pe}U_m(2zc_2-z^2c_1),
\end{equation}
\begin{equation}
c_j(0)=c_j(\infty)=0, \;\; \mathrm{for} \;\; j\geq0, \quad
c_j'(0)=\begin{cases}
-8/V^2, \;\; \mathrm{for} \;\; j=0,\\
0, \;\; \mathrm{otherwise}.
\end{cases}
\end{equation}
\end{subequations}
The above equations can be directly solved with the first two boundary conditions. Using the third boundary condition, one can then calculate the growth rate as
\begin{equation}\label{eq:largek1}
\sigma=\sigma_{-2}k^2+\sigma_{0}+\frac{\sigma_{1}}{k}+\frac{\sigma_{2}}{k^2}+\cdots,
\end{equation}
with
\begin{subequations}
\begin{equation}\label{eq:largek1a}
\sigma_{-2}=\frac{\mathrm{Pe}}{8}V^2-\sqrt{\frac{\mathrm{Pe}}{2}}V,
\end{equation}
\begin{equation}\label{eq:largek1b}
\sigma_0=\frac{i\mathrm{Pe}U_m}{\sigma_{-2}\sigma_*}
(3\sigma_*-3-4\sigma_{-2}),
\end{equation}
\begin{equation}
\sigma_1=\frac{i\mathrm{Pe}U_m}{2\sigma_0^2\sigma_*^3}
\left[
16(\sigma_*-1)+\sigma_0(27\sigma_*-35)+\sigma_0^2(12\sigma_*-19)
\right],
\end{equation}
\begin{equation}
\sigma_2=\frac{\mathrm{Pe}^2U_m^2}{4\sigma_{-2}^3\sigma_*^5}
\left[60(1-\sigma_*)+\sigma_{-2}(179-149\sigma_*)
+\sigma_{-2}^2(178-111\sigma_*)+\sigma_{-2}^3(59-24\sigma_*)\right],
\end{equation}
\end{subequations}
where $\sigma_{-2}$ recovers the result without the imposed flow \cite{rubinstein2000electro} and $\sigma_*=\sqrt{1+\sigma_{-2}}$. In the presence of the pressure-driven flow, a perturbation of large wavenumber propagates with an O$(1/k)$ wave speed and its growth rate is decreased by an O$(1/k^2)$ term. The leading order ion concentration is
\begin{equation}\label{eq:largek1c}
c_0=-\frac{\mathrm{Pe}}{\sigma_{-2}^2}
(2e^{-\sqrt{1+\sigma_{-2}}z}
+(\sigma_{-2}z-2)e^{-z}).
\end{equation}
The imposed flow reduces the growth rate of a perturbation at high wavenumbers and the stabilizing effect becomes less effective as $k$ increases. As we will see in the following, this result is consistent with the full analysis. The bulk analysis predicts that a full suppression of electroconvection by an imposed cross-flow is impossible. This  result is qualitatively different from what we will obtain from the full analysis.

\subsubsection{Numerical results for all wavenumbers}
\begin{figure}
\begin{center}
\includegraphics[angle=0,scale=0.34]{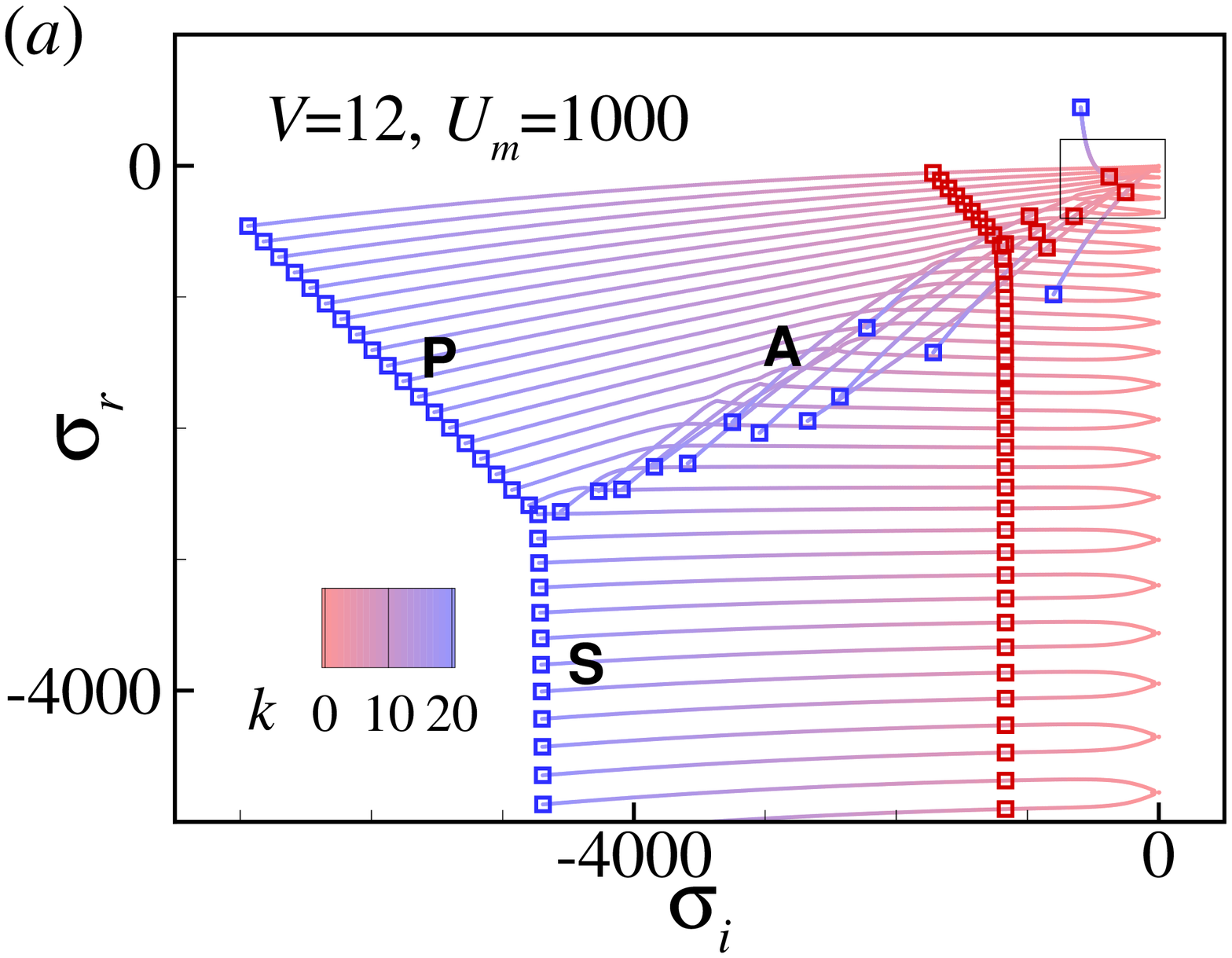}
\includegraphics[angle=0,scale=0.34]{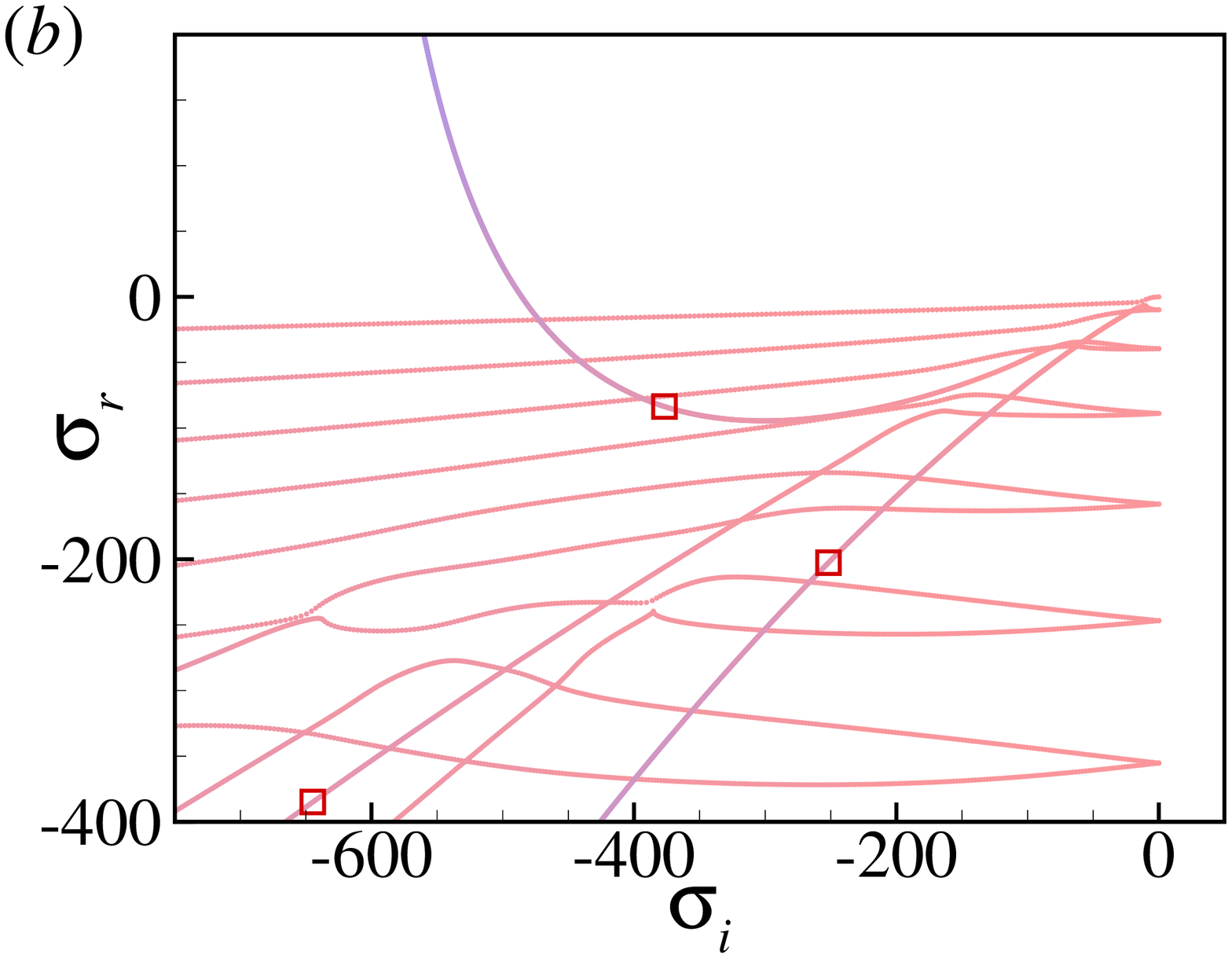}
\caption{($a$) The spectrum of the purely electroconvective instability in the range $0\leq k\leq20$ at $V=12$ and $U_m=1000$. The squares highlight the eigenvalues at two wavenumbers $k=5$ (red) and $k=20$ (blue). ($b$) is a closer view of ($a$). }\label{fig:fig3}
\end{center}
\end{figure}

For arbitrary wavenumber, we apply a Chebyshev collocation method ~\cite[]{weideman2000matlab} to numerically solve equ. (\ref{eq:perturbed}) and (\ref{eq:perturbedbc1}) for the eigensolutions. Figure \ref{fig:fig3}($a$) shows the eigenvalue spectrum for the purely electroconvective instability for $0\leq k\leq20$, $V=12$, and $U_m=1000$ which corresponds to a velocity of $470\mu\mathrm{m}/\mathrm{s}$ in a channel of 2mm width. As in the high-Reynolds number channel flow without electrokinetic effects, the spectrum has three branches of solutions. The A-branch solution of small wave speed corresponds to the wall mode, the P-branch of large wave speed corresponds to the center mode and the S-branch is the highly damped mode whose wave speed is the average velocity $2\mathrm{Pe}U_m/3$. The P and S branches have very similar structures to those in high-Reynolds number channel flow without electrokinetic effects ~\cite[]{schmid2012stability}, indicating that the P and S modes are mainly caused by the parabolic velocity profile of the Poiseuille flow. The A-branch is the mode which causes the electroconvective instability.  Fig. \ref{fig:fig3}($b$) shows a closer view of the spectrum. At small wavenumber, all the modes are stable and the largest growth rate of the perturbations belongs to the P-branch solution. As the wavenumber increases, the growth rate of the center mode decreases while that of the wall mode increases eventually resulting in an instability. This type of instability caused by the electroosmotic slip velocity is qualitatively different from the one in a channel flow due to inertial effects.

\begin{figure}
\begin{center}
\includegraphics[angle=0,scale=0.34]{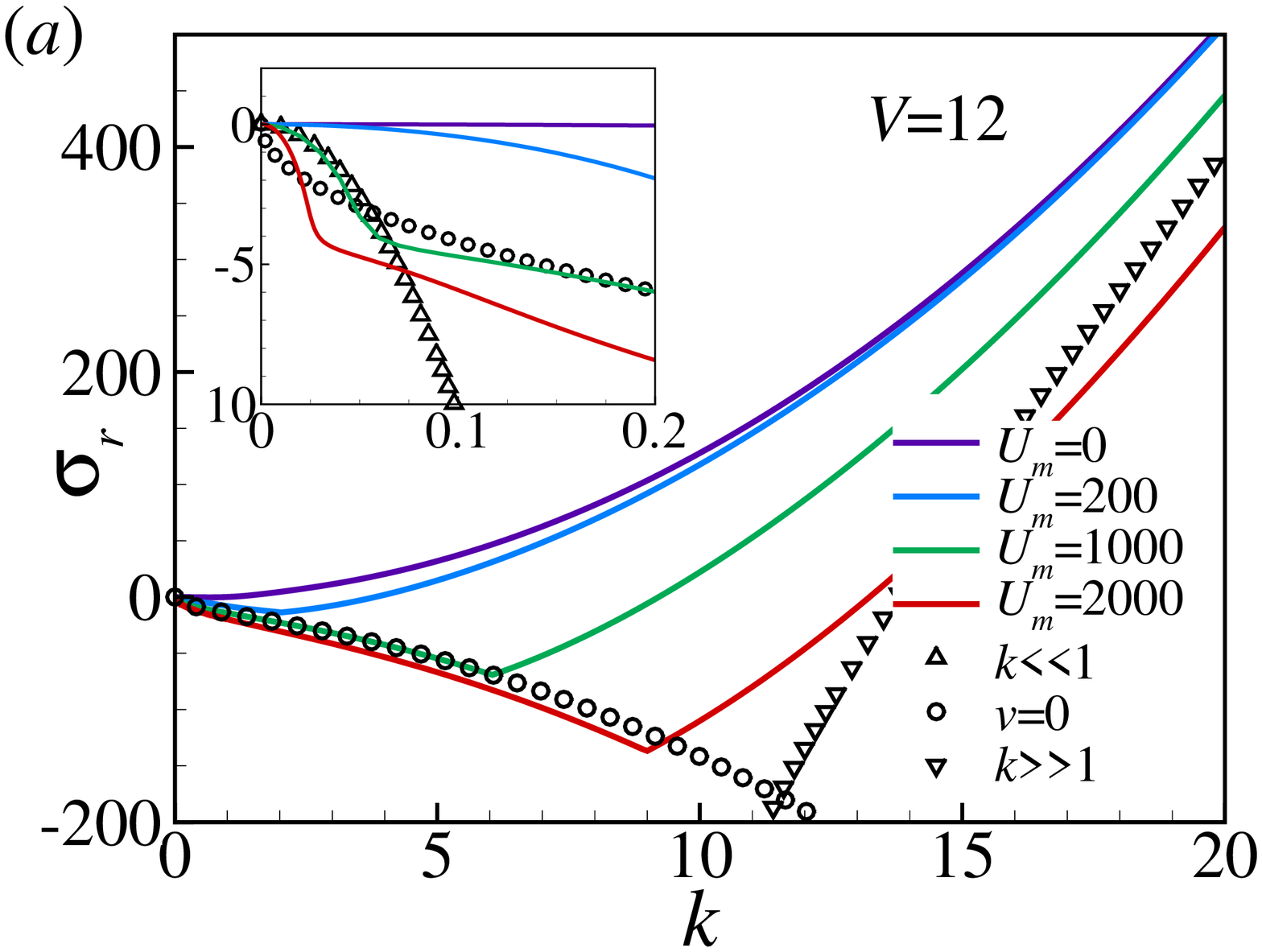}
\includegraphics[angle=0,scale=0.34]{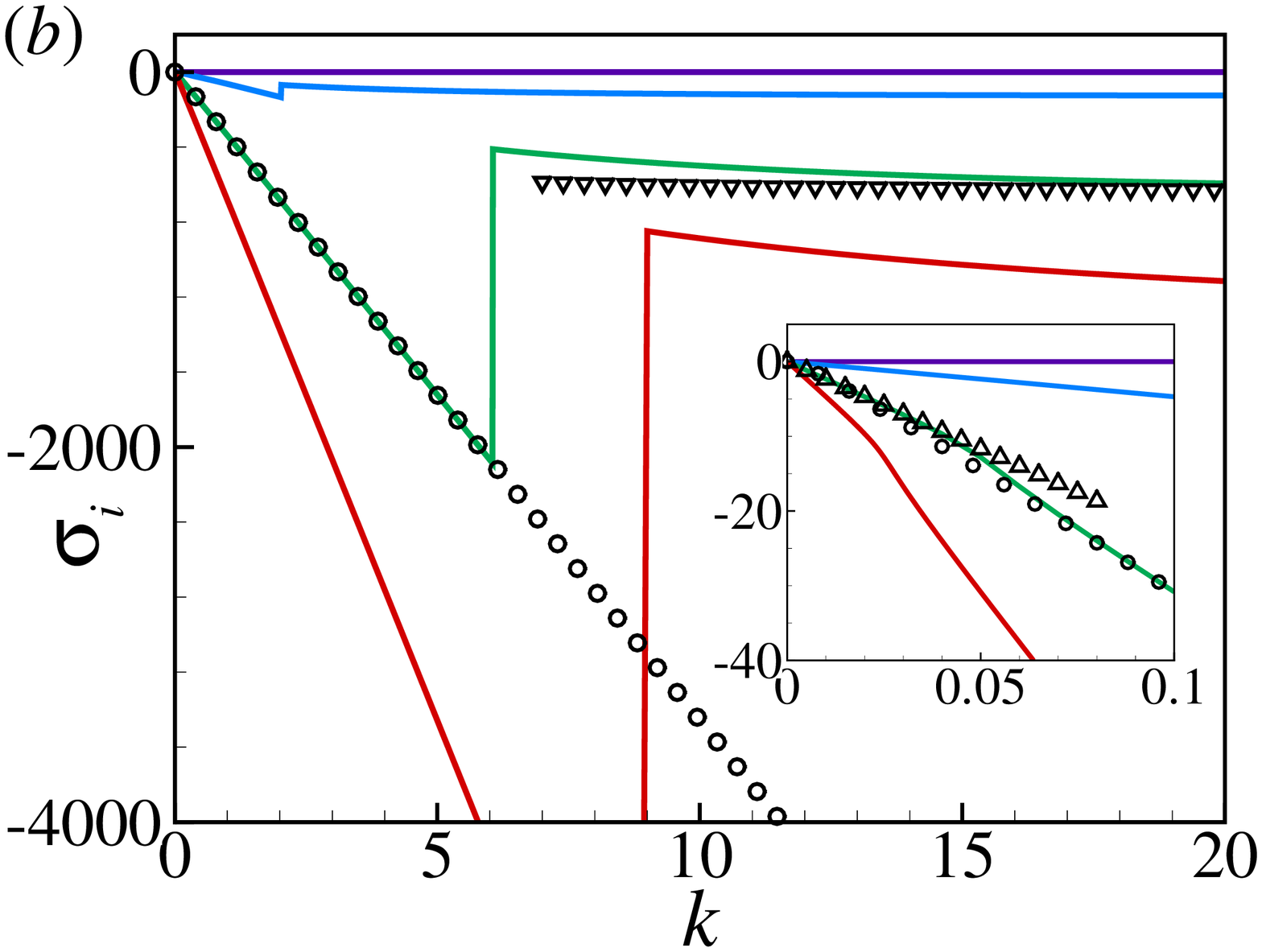}
\caption{The complex growth rate ($a$: real, $b$: imaginary) of the most unstable mode for the purely electroconvective instability at different velocities at $V=12$. The asymptotic solutions for $U_m=1000$ are derived in equations (\ref{eq:smallk1}), (\ref{eq:vequ0}) and (\ref{eq:largek1}) for $k\ll1$, $v=0$ and $k\gg1$, respectively.}\label{fig:fig4}
\end{center}
\end{figure}

We now focus on the mode with the largest growth rate, which determines the linear instability of the flow. Figure \ref{fig:fig4} shows the eigenvalues of the most unstable mode as a function of the wavenumber $k$ at $V=12$. The symbols show the asymptotic solutions for $U_m=1000$. For both real and imaginary parts of the growth rate, the abrupt changes in the curves manifest the transition from the centerline mode to the wall mode with increasing $k$, and the transition wavenumber $k_{tr}$ increases with increasing velocity of the imposed flow. The centerline mode is always stable, while the wall mode eventually becomes unstable at large $k$, showing that the imposed flow can only suppress the electroconvection at large length scales. This effect is directly caused by the relative importance of the imposed flow and the second-kind electroosmotic slip velocity which causes the electroconvective instability. The second-kind slip velocity is generated by the tangential gradient of the perturbed ion concentration ~\cite[]{rubinstein2005electroconvective} and is proportional to $k^2$. At small wavenumber, the slip velocity is negligible compared to the cross-flow and the linear instability is determined by the laminar channel flow. Therefore, the modes decay with $\sigma_r\sim-k^2$ due to the diffusion of the perturbed ion concentration. At large enough wavenumbers, the destabilizing effect of the slip velocity on the concentration field can always overtake the stabilizing effect due to the imposed flow and therefore leads to the electroconvective instability. In figure \ref{fig:fig4}($b$), the centerline mode perturbation propagates downstream with a constant wave speed $u_c=-\sigma_i/k=PeU_m$ which is exactly the velocity of the imposed flow at the centerline. The wall mode perturbation has a smaller wave speed $u_c\sim1/k$, indicating that near the wall the small vortices propagate slower than the large vortices.

\begin{figure}
\begin{center}
\includegraphics[angle=0,scale=0.34]{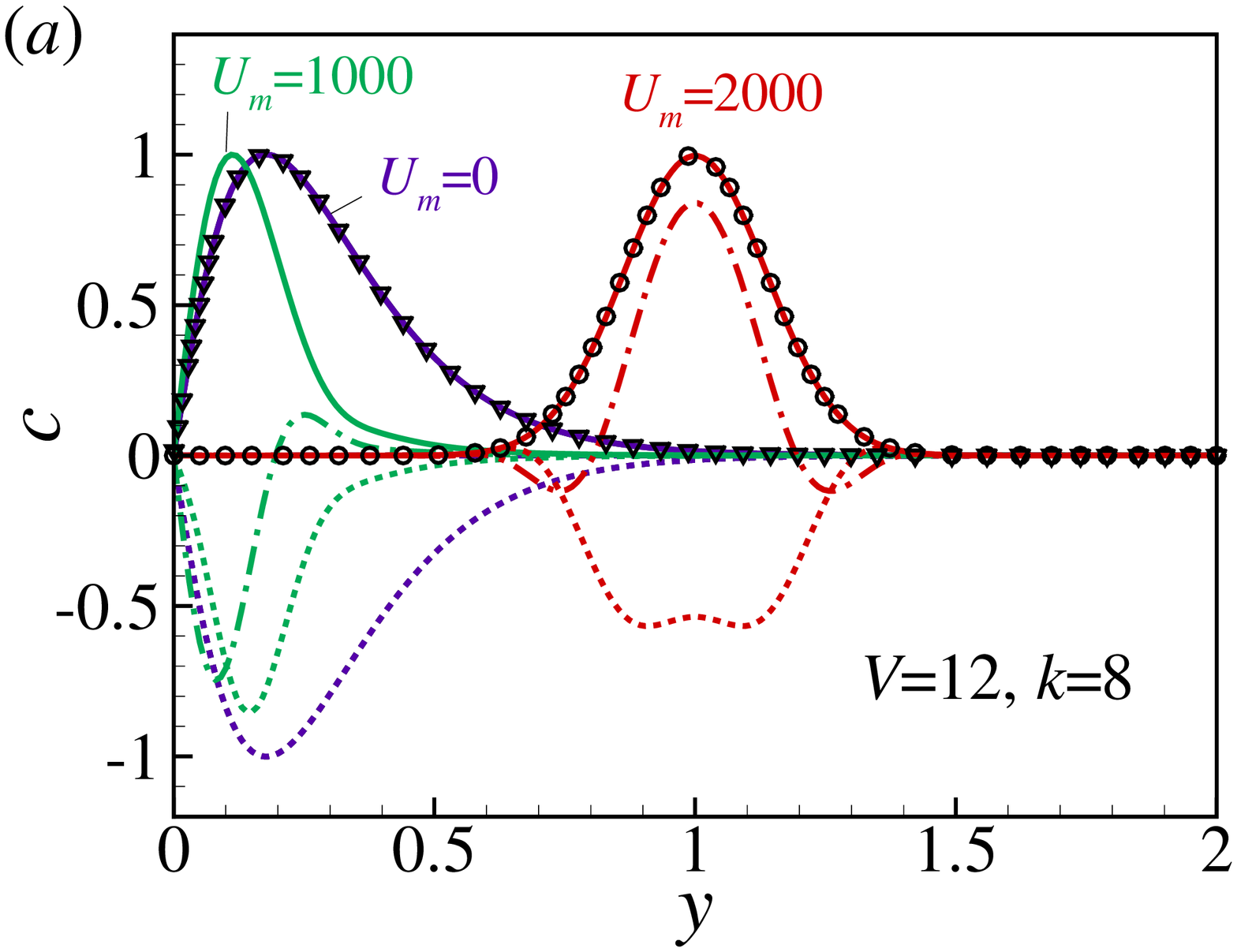}
\includegraphics[angle=0,scale=0.34]{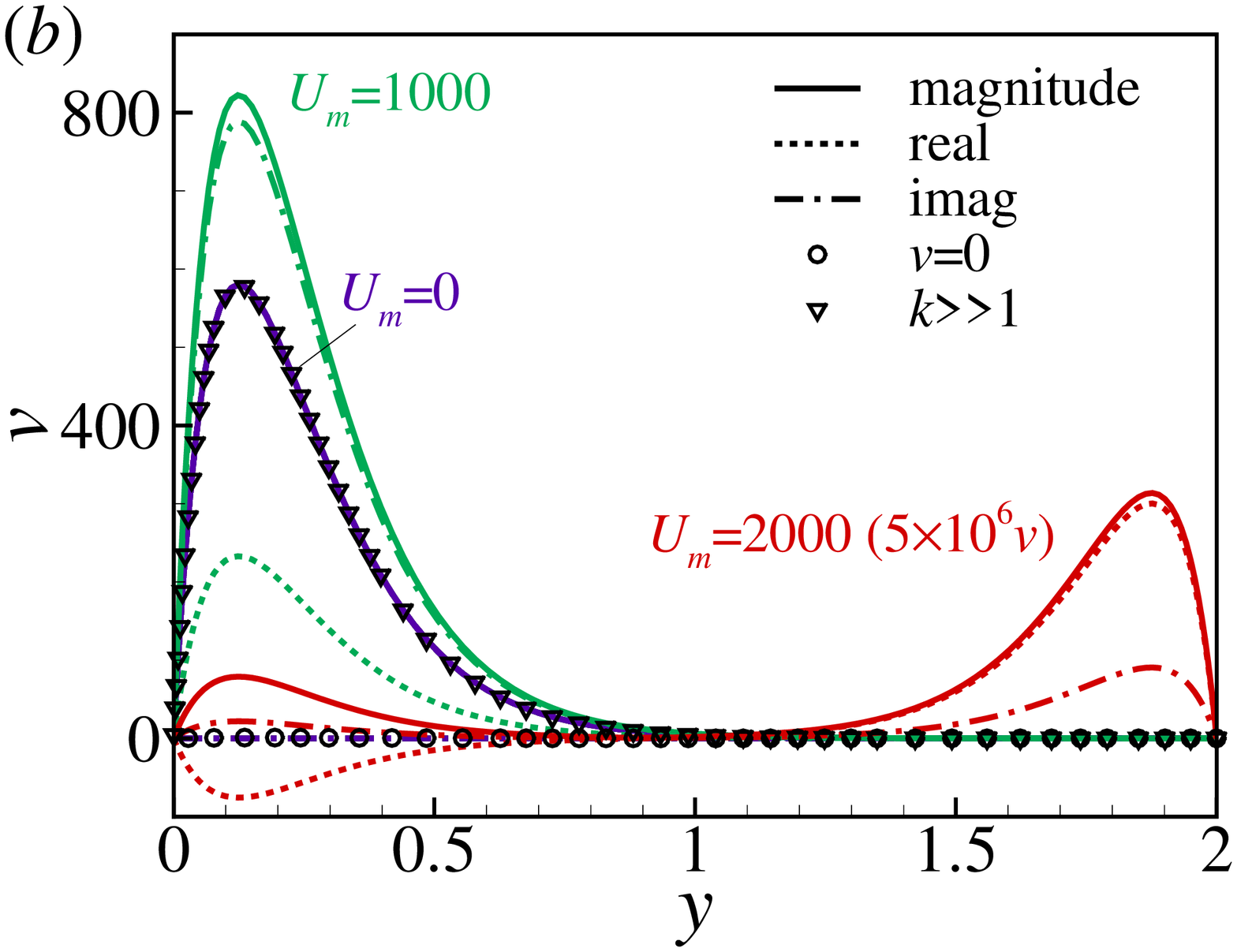}
\caption{Distribution of the perturbed ($a$) ion concentration and ($b$) normal velocity of the most unstable mode for the electroconvective instability at $U_m=0, 1000$ and 2000, for $V=12$ and $k=8$. Perturbations grow at $U_m=0$ and decay at $U=1000$ and 2000. The velocity for $U_m=2000$ is multiplied by $5\times10^6$. Symbols show the magnitude of the analytical solutions for $v=0$ and $k\gg1$ given in Equ. (\ref{eq:vequ0c}) and (\ref{eq:largek1v}), (\ref{eq:largek1c}).}\label{fig:fig5}
\end{center}
\end{figure}

\begin{figure}
\begin{center}
\includegraphics[angle=0,scale=0.32]{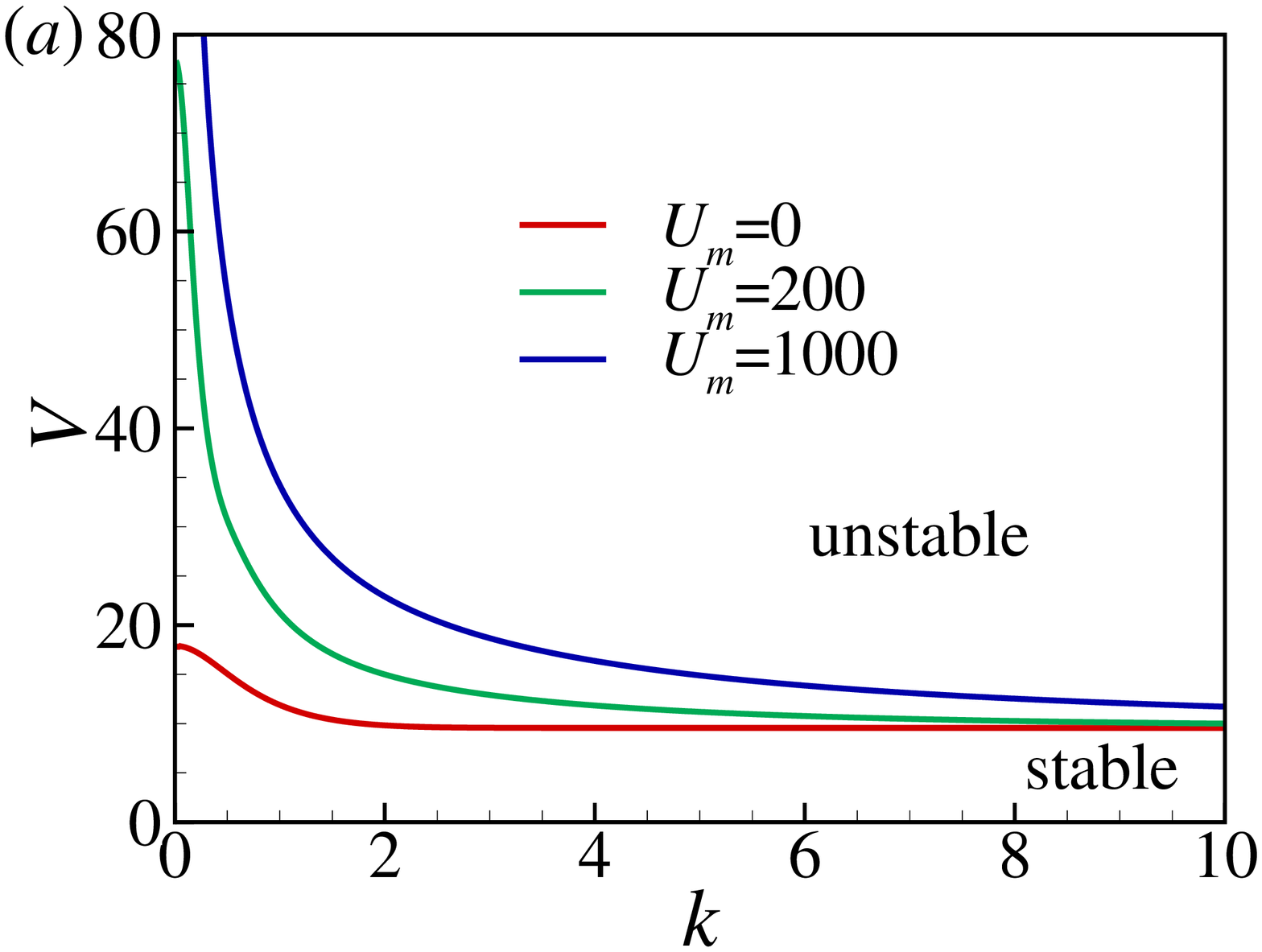}
\includegraphics[angle=0,scale=0.32]{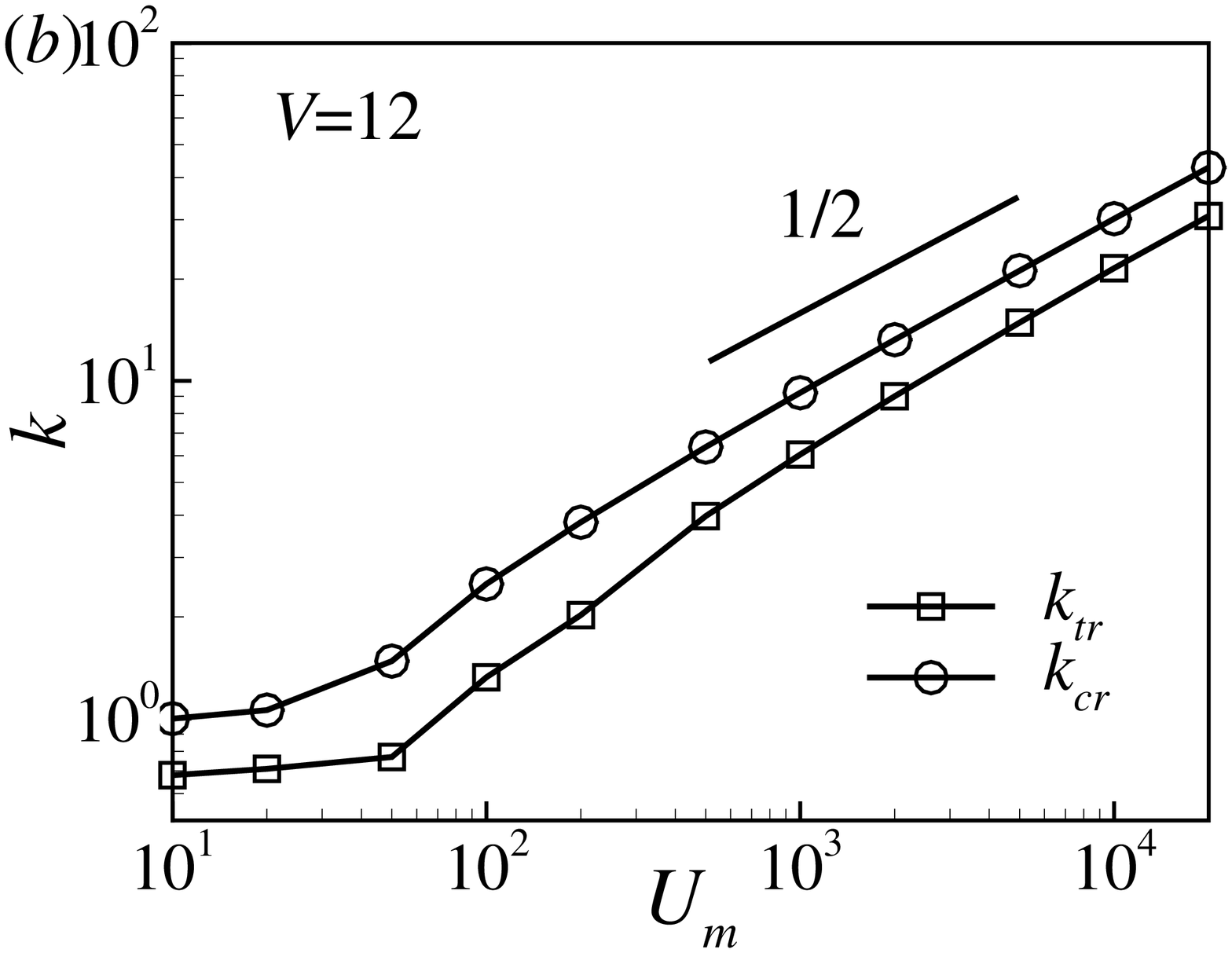}
\caption{($a$) The marginal stability curves in the $k-V$ space at different cross-flow velocities. ($b$) The dependence of $k_{cr}$ and $k_{tr}$ on $U_m$ at a constant potential $V=12$. $k_{cr}$ is the critical wavenumber at which the mode becomes unstable, $k_{tr}$ is the transition wavenumber between centerline and wall modes and has the minimum growth rate.}\label{fig:fig6}
\end{center}
\end{figure}

To examine the manner in which modes are stabilized by the imposed flow, we compare the eigenfunctions of the most dangerous mode at different velocities fig. \ref{fig:fig5}. The results are normalized such that the magnitude of the ion concentration has the same maximum value $c_{max}=1$. At $V=12$ and $k=8$, the perturbations grow at $U_m=0$ and decay at $U_m=1000$ and 2000. At a high enough flow rate $U_m=2000$, the perturbed ion concentration has a Gaussian distribution at the centerline of the channel. The ion concentration gradients at the two walls are small and therefore the induced electroosmotic slip velocities are negligible on both sides. Note that the velocity for $U_m=2000$ has been multiplied by $5\times10^6$ for clarity.
In comparison, at $U_m=0$ the ion concentration has a large perturbation at the anode surface and induces a strong slip velocity at the bottom of the channel.
The real components of $c$ and $v$ have opposite signs near $y=0$. This means that the increase of the ion concentration generates a local downward flow which flushes more ions into this region and further amplifies the perturbation.
At $U_m=1000$, the perturbations are concentrated near the bottom surface. The magnitude of the velocity $v$ is higher than in the case with $U_m=0$, while the real part is smaller, showing that the imposed flow suppresses the instabilities by reducing the coupling between the electroosmotic slip velocity and the ion concentration gradient.

The marginal stability curves in the $k-V$ space at different imposed velocities are compared in figure \ref{fig:fig6}($a$). At $U_m=0$, the critical voltage for the pure electroconvective instability is $V_{cr1}=\sqrt{48(1+1/D)/\mathrm{Pe}-32\ln2}=14.4$ for $k\ll1$ and $V_{cr2}=\sqrt{32/Pe}=9.6$ for $k\gg1$. As shown in the previous results, the imposed flow only increases the critical voltage at small wavenumber, while at large $k$, all the curves reach the same critical  voltage. At a fixed voltage, increasing the velocity $U_m$ increases the critical wavenumber $k_{cr}$ at which the mode becomes unstable. In Fig. \ref{fig:fig6}($b$), both the critical wavenumber $k_{cr}$ and the transition wavenumber between the two modes $k_{tr}$ increase as $U_m^{1/2}$ at large $U_m$, suggesting that the electroconvection is stabilized when the concentration field is more influenced by the imposed flow $U_m$ than the electroosmotic slip velocity $u_s\sim k^2$. To summarize, the bulk analysis predicts that the pressure-driven flow can attenuate the small wavenumber modes that would arise in a pure electroconvective instability, and the critical wavenumber above which modes are suppressed increases with flow speed as $k_{cr}\sim U_m^{1/2}$.

\subsection{Coupled electroconvective and morphological instability}
In this subsection, we consider the coupled electroconvective and morphological instability. In contrast to the pure electroconvective instability, the coupled electroconvective and morphological instability is always unstable. In the following, we will first discuss the asymptotic solutions for $k\ll1$ and $k\gg1$, and then show the numerical result for arbitrary wavenumber.

\subsubsection{Small wavenumber, $k\ll1$}
For $k\ll1$, the variables are expanded as $c=c_0+kc_1+\cdots, \mu=\mu_0+k\mu_1+\cdots, v=v_0+kv_1+\cdots,$ and $\sigma=\sigma_0+k\sigma_1+\cdots$. Different from the purely electroconvective instability, here the velocity $v$ is directly caused by the growth of the electrode surface and its leading term is O(1) instead of O($k^2$). Therefore, the two highest order equations are $v_j^{(4)}=0$ with boundary conditions $v_j(0)=\sigma_ih/\mathrm{Pe}$ and $v_j'(0)=v_j(2)=v_j'(2)=0$ for $j=0,1$. The solutions are
\begin{equation}
v_j=\frac{(y+1)(y-2)^2}{4\mathrm{Pe}}\sigma_jh,
\;\;\mathrm{for}\;\;j=0,1.
\end{equation}
For the ion concentration, the leading order equation and the boundary conditions are
\begin{subequations}
\begin{equation}\label{eq:cmc0}
\sigma_0c_0+\mathrm{Pe}v_0=c_0'',
\end{equation}
\begin{equation}
c_0(0)+h=0, \quad
c_0'(0)=\frac{\sigma_0h}{(1+D)v_m},
\end{equation}
\end{subequations}
leading to the solution
\begin{equation}
\begin{split}
c_0=& \left(\frac{\sqrt{\sigma_0}h}{(1+D)v_m}
+\frac{3h}{2\sqrt{\sigma_0^3}}\right)\sinh(\sqrt{\sigma_0}y) \\
- & \frac{h}{4\sigma_0}\left(6\cosh(\sqrt{\sigma_0}y)
+6(y-1)+(y+1)(y-2)^2\sigma_0\right).
\end{split}
\end{equation}
Combining the equations for ion concentration and chemical potential, and using the boundary conditions $\mu_0'(0)=\mu_0'(2)=0$, one can derive $c_0'(2)=c_0'(0)$ and find the leading order growth rate at $k=0$,
\begin{equation}\label{eq:puremorph0}
\sqrt{\sigma_0}\coth\sqrt{\sigma_0}=
1+\frac{2\sigma_0^2}{3(1+D)v_m},
\end{equation}
where $D=D^+/D^-$ is the ratio of the cation and anion diffusivities, and $v_m=0.013$ is the dimensionless molar volume of the lithium metal. In typical electrolytes, $D\sim1$ and $v_m\ll1$, the growth rate $\sigma_0\ll1$. Using the approximation $\sqrt{\sigma_0}\coth\sqrt{\sigma_0}\simeq1+\sigma_0/3$, we get $\sigma_0\simeq(D+1)v_m/2$ for $v_m\ll1$, and the ion concentration $c_0\simeq h(y/2-1)$.

As a comparison, the growth rate for the purely electroconvective instability is $\sigma=0$ at $k=0$. On the other hand, the growth rate for the purely morphological instability follows
\begin{equation}\label{eq:puremorph1}
\coth\sqrt{\sigma_0}=\frac{\sqrt{\sigma_0}}{(1+D)v_m},
\end{equation}
and the approximate solution is $\sigma_0\simeq(D+1)v_m$ and $c\simeq h(y-1)$ for $v_m\ll1$. The morphological instability is intrinsically unstable even at $k=0$. This result can be understood from the boundary condition at the electrode surface. Consider a local peak on the anode surface, the boundary condition $c(0)+h=0$ reduces the ion concentration at the surface and enhances the ion flux by increasing the local concentration gradient, thereby causing more deposition. Allowing the electroconvective effect smooths the ion concentration gradient at $y=0$, and reduces the growth rate by half. This is the leading order growth rate of the perturbation, which is not affected by the pressure-driven flow.

%Since the leading order growth rate is unaffected by the pressure-driven flow, we continue to solve the growth rate on the next order. The governing equation for the ion concentration on O($k$) is
%\begin{subequations}\label{eq:cmc1}
%\begin{equation}
%\sigma_0c_1+\sigma_1c_0+iPeU_my(2-y)c_0+\mathrm{Pe}v_1=c_1'',
%\end{equation}
%\begin{equation}
%c_1(0)=0, \quad
%c_1'(0)=\frac{\sigma_1h}{(1+D)v^*_m}.
%\end{equation}
%\end{subequations}
%Solve the above equation for $c_1$ and substitute equ. (\ref{eq:cmc1}) into next order equation for $\mu_1$
%\begin{equation}
%\sigma_0c_1+\sigma_1c_0+iPeU_my(2-y)c_0+\mathrm{Pe}v_1=\frac{D+1}{2D}(y\mu_1')',
%\end{equation}
%and the boundary conditions $\mu_1'(0)=0$ and $\mu_1'(2)=0$, one can get $c_1'(2)-c_1'(0)=0$ to solve the next order growth rate $\sigma_1$.

The pressure-driven flow affects the eigenvalue at higher orders. Here we only consider the next order solution with $v_m\ll 1$, for which the eigenvalue
\begin{equation}\label{eq:smallk2}
\sigma_1\simeq2i\mathrm{Pe}U_m/3,
\end{equation}
is purely imaginary. Like the purely electroconvective mode, the morphological perturbation at small $k$ has a wave speed equal to the average fluid velocity. However, its propagation is in the opposite direction, against the imposed flow. This is because the cross-flow brings more ions to the windward side of a perturbed surface and increases the local deposition on this side, and therefore causes the wave to propagate upstream.
Since the applied flow suppresses the electroconvection, we expect the pressure-driven flow to increase the growth rate for the coupled electroconvective and morphological instability for $k\ll1$.

%\subsubsection{small molar volume limit}
%We expand the equations first in terms of molar volume $v^*_m$ and then expand each term on $k$, i.e., $c=(c_{-1,0}+c_{-1,1}k+\cdots)/v^*_m+(c_{0,0}+c_{0,1}k+\cdots)+\cdots, v=(v_{-1,2}k^2+v_{-1,3}k^3+\cdots)/v^*_m+(v_{0,2}k^2+v_{0,3}k^3+\cdots)+\cdots, \mu=(\mu_{-1,0}+\mu_{-1,1}k+\cdots)/v^*_m+(\mu_{0,0}+\mu_{0,1}k+\cdots)+\cdots$ and $\sigma=(\sigma_{0,0}+\sigma_{0,1}k+\cdots)+(\sigma_{1,0}+\sigma_{1,1}k+\cdots)v^*_m+\cdots$. The perturbed equations at the first two leading orders are
%\begin{subequations}
%\begin{equation}
%c_{-1,0}=\sin(\sigma_{0,0}y)
%\end{equation}
%\begin{equation}
%\sigma_{0,0}=-(n\pi)^2
%\end{equation}
%\begin{equation}
%c_{-1,1}=\frac{\sigma_{0,1}h}{D+1}
%\end{equation}
%\begin{equation}
%\sigma_{0,1}=-\frac{2i\mathrm{Pe}U_m}{3}
%\end{equation}
%\end{subequations}

\subsubsection{Large wavenumber, $k\gg1$}
For $k\gg1$, the analysis is performed in the vicinity of the depletion anode surface with an inner scale $z=ky$ and the electrolyte is considered to be semi-infinite. In our previous study on the bulk region ~\cite[]{tikekar2018electroconvection}, we derived the analytical solution for $k\gg1$ without a flow. The growth rate monotonically increases with the wavenumber $k$, and its scaling depends on the applied voltage. For $V<V_{\mathrm{cr2}}$, where $V_{\mathrm{cr2}}=\sqrt{32/\mathrm{Pe}}$ is the critical voltage for the onset of electroconvection for $k\gg1$, the growth rate $\sigma\sim k$. For $V>V_{\mathrm{cr2}}$, the instability is mainly contributed by the electroconvective instability and $\sigma\sim k^2$. The growth rate for the coupled instability in an imposed flow has the same scalings.

For $V<V_{\mathrm{cr2}}=\sqrt{32/\mathrm{Pe}}$, we have
\begin{equation}
\sigma=k\sigma_{-1}+\sigma_0+\frac{\sigma_1}{k}+\cdots,
\end{equation}
with
\begin{subequations}
\begin{equation}
\sigma_{-1}=\frac{(D+1)v_m}{1-PeV^2/32}.
\end{equation}
\begin{equation}
\sigma_0=-256(D+1)^2v_m^2
\frac{1+\mathrm{Pe}V^2/32}{(1-\mathrm{Pe}V^2/32)^3}.
\end{equation}
\begin{equation}\label{eq:largek2c}
\sigma_1=\frac{(D+1)^3v_m^3}{16}
\frac{3+5\mathrm{Pe}V^2/32}{(1-\mathrm{Pe}V^2/32)^5}
+\frac{i(D+1)v_m\mathrm{Pe}U_m}{4}
\frac{2-7\mathrm{Pe}V^2/32}{(1-\mathrm{Pe}V^2/32)^2}.
\end{equation}
\end{subequations}
The imposed flow reduces the growth rate of the coupled instability at high wavenumbers. The wave speed of the perturbation scales as $u_c=\sigma/k\sim\mathrm{O}(k^{-2})$, the wave can propagate either downstream or upstream depending on the applied voltage.

The leading order solutions for the ion concentration and normal velocity are
\begin{subequations}\label{eq:largek2a}
\begin{equation}
c=-he^{-z}+\frac{\mathrm{Pe}V^2h}{32-PeV^2}z(z+1)e^{-z},
\end{equation}
\begin{equation}
v=-\frac{k^2h}{8-\mathrm{Pe}V^2/4}ze^{-z}.
\end{equation}
\end{subequations}
Note that the normal velocity boundary condition $v(0)=\sigma h/\mathrm{Pe}$ does not affect the leading order equation since $\sigma\sim k$ has a smaller order than the normal velocity $v\sim k^2$. In other words, the growth of the anode surface makes a negligible contribution to the fluid velocity. Instead, the flow is mainly induced by a tangential electroosmotic slip velocity (\ref{eq:perturbedbc1d}) due to the ion concentration gradient caused by the surface perturbation.

As a comparison, the growth rate for the purely morphological instability at $k\gg1$ is ~\cite[]{tikekar2018electroconvection}
\begin{equation}\label{eq:puremorph2}
\sigma=(D+1)v_mk
\end{equation}
and the perturbed ion concentration is
\begin{equation}
c=-he^{-z}.
\end{equation}
At large wavenumber, the electroosmotic slip flow increases the growth rate of the morphological instability, even before the onset of electroconvection. This is because the electroosmotic flow near the electrode surface increases the ion concentration gradient and therefore enhances the instability. This result is opposite to the growth rate at small wavenumber, for which the electroconvection reduces the growth rate of the morphological instability.

%The leading order ion concentration and velocity are
%\begin{equation}
%c=\frac{\mathrm{Pe}h}{32-\mathrm{Pe}V^2}ky(ky+1)e^{-ky}-he^{-ky}+\cdots,
%\end{equation}
%and
%\begin{equation}
%v=-\frac{k^3h}{8-\mathrm{Pe}V^2/4}ye^{-ky}+\cdots.
%\end{equation}

%\begin{equation}
%v_{-2}=ze^{-z},
%\end{equation}
%due to the homogenous boundary condition $v_{-2}(0)=0$. For the ion concentration, the leading order equation on O(1) is
%\begin{equation}
%c_0''-c_0=\mathrm{Pe}v_{-2},
%\end{equation}
%with three boundary conditions at $y=0$
%\begin{equation}
%c_0(0)+h=0, \quad
%c_0'(0)=-8v_{-2}'(0)/V^2, \quad
%c_0'(0)=\frac{\sigma_{-1}}{(D+1)v^*_m}h,
%\end{equation}
%and the bound condition $c_0(\infty)=0$ at infinity. Two of the conditions determine the concentration profile
%\begin{equation}
%c_{0}=-\frac{\mathrm{Pe}}{4}z(z+1)e^{-z}-he^{-z},
%\end{equation}
%and the other two conditions determine the height of the interface
%\begin{equation}
%h=\frac{\mathrm{Pe}}{4}-\frac{8}{V^2},
%\end{equation}
%and the growth rate
%\begin{equation}
%\sigma_{-1}=\frac{(D+1)v^*_m}{1-PeV^2/32}.
%\end{equation}

For $V>V_{\mathrm{cr2}}$, the instability is dominated by electroconvection and the growth rate is
\begin{equation}
\sigma=k^2\sigma_{-2}+\cdots,
\end{equation}
with $\sigma_{-2}=\mathrm{Pe}V^2/8-\sqrt{\mathrm{Pe}V^2/2}$ which is the same as for the purely electroconvective instability in equ. (\ref{eq:largek1a}).

The leading order solutions for ion concentration and velocity are
\begin{subequations}\label{eq:largek2b}
\begin{equation}
c=\frac{\mathrm{Pe}V^2kh}{8(D+1)v_m\sigma_{-2}}\left(
2e^{-\sqrt{\sigma_{-2}+1}z}+(\sigma_{-2}z-2)e^{-z}\right),
\end{equation}
\begin{equation}
v=-\frac{\sigma_{-2}V^2k^3h}{8(D+1)v_m}ze^{-z},
\end{equation}
\end{subequations}
which are the same as equ. (\ref{eq:largek1v}) and (\ref{eq:largek1c}) for the purely electroconvective instability.

%It can be found that the growth rate satisfies
%\begin{equation}
%\sigma=\frac{v^*_mk_c}
%{\frac{1}{D+1}-\frac{\mathrm{Pe}V^2}{8(D+1)}\frac{k^2}{(k+k_c)^2}
%+v^*_m\frac{2k+k_c}{(k+k_c)^2}},
%\end{equation}
%where $k_c=\sqrt{k^2+\sigma}$

%\begin{equation}
%v=\frac{\sigma h}{\mathrm{Pe}}e^{-ky}
%+(\frac{\sigma kh}{\mathrm{Pe}}-\frac{V^2k^2}{8}c'(0))ye^{-ky}.
%\end{equation}
%For the ion concentration, the equation is
%\begin{equation}
%c''-(k^2+\sigma)c=\mathrm{Pe}v,
%\end{equation}
%with boundary conditions
%\begin{equation}
%c(0)+h=0, \quad
%c'(0)=\frac{\sigma}{(D+1)v^*_m}h, \quad
%c(\infty)=0.
%\end{equation}

\subsubsection{Numerical results for all wavenumbers}

\begin{figure}
\begin{center}
\includegraphics[angle=0,scale=0.32]{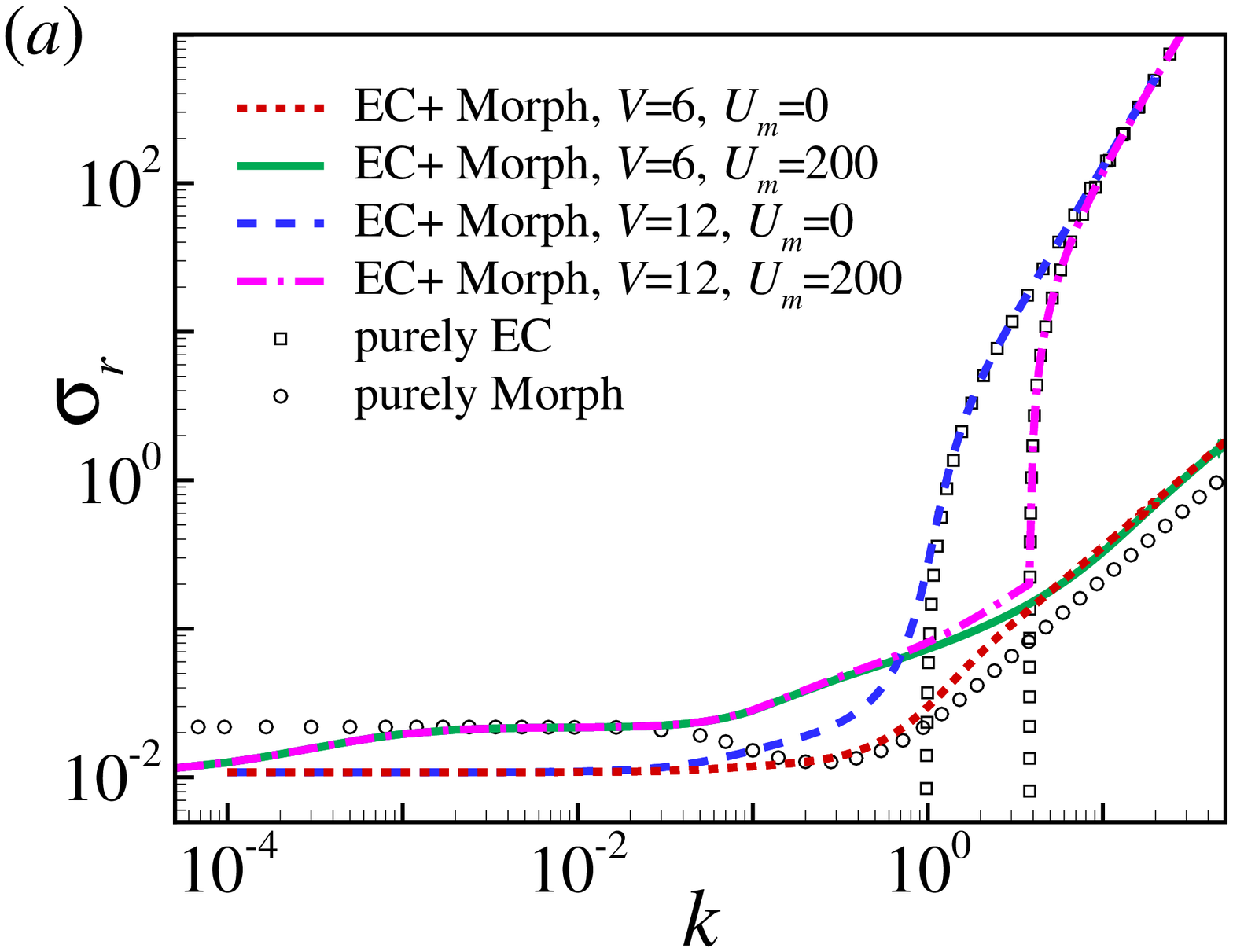}
\includegraphics[angle=0,scale=0.32]{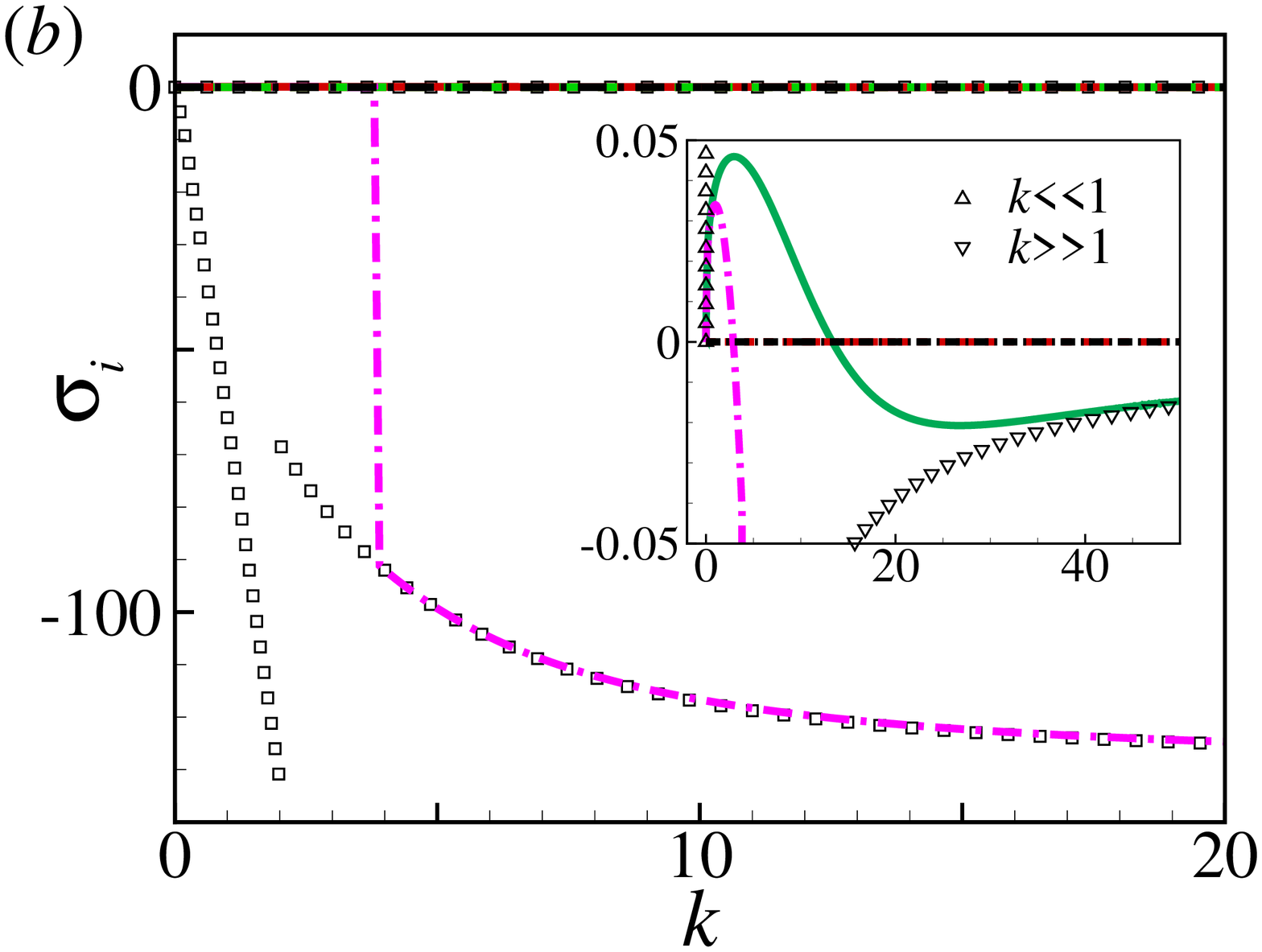}
\caption{The maximum growth rate ($a$: real, $b$: imaginary) of the electroconvective, morphological and the coupled instabilities with and without the cross-flow. The asymptotic solutions of $\sigma_i$ for $k\ll1$ and $k\gg1$ are given in Equ. (\ref{eq:smallk2}) and (\ref{eq:largek2c}), respectively.}\label{fig:fig7}
\end{center}
\end{figure}

Figure \ref{fig:fig7}($a$) shows the growth rate of the most unstable mode for the purely electroconvective instability, the purely morphological instability and the coupled instability. For the morphological instability, the growth rate is independent of the applied potential since the base state current is a constant $I=2$. Its growth rate scales as $\sigma_r\sim(D+1)v_m$ for $k\ll1$ and $\sigma_r\sim(D+1)v_mk$ for $k\gg1$, and it has a minimum growth rate at $k\sim0.3$. Allowing the electroconvective instability decreases the growth rate for small $k$ while increasing it for large $k$. The pressure-driven flow reduces the effects of electroconvection, it increases the growth rate of the perturbation for small $k$, and decreases the growth rate for $k>1$ but eventually its effect becomes negligible. In Figure \ref{fig:fig7}($b$), the perturbation becomes a traveling wave even without the onset of the electroconvection due to a nonuniform deposition on the upwind and downwind sides of the perturbation, and it propagates upstream at small $k$ and downstream at large $k$. The electroconvection substantially increases the wave speed of the perturbation at large $k$.

\begin{figure}
\begin{center}
\includegraphics[angle=0,scale=0.32]{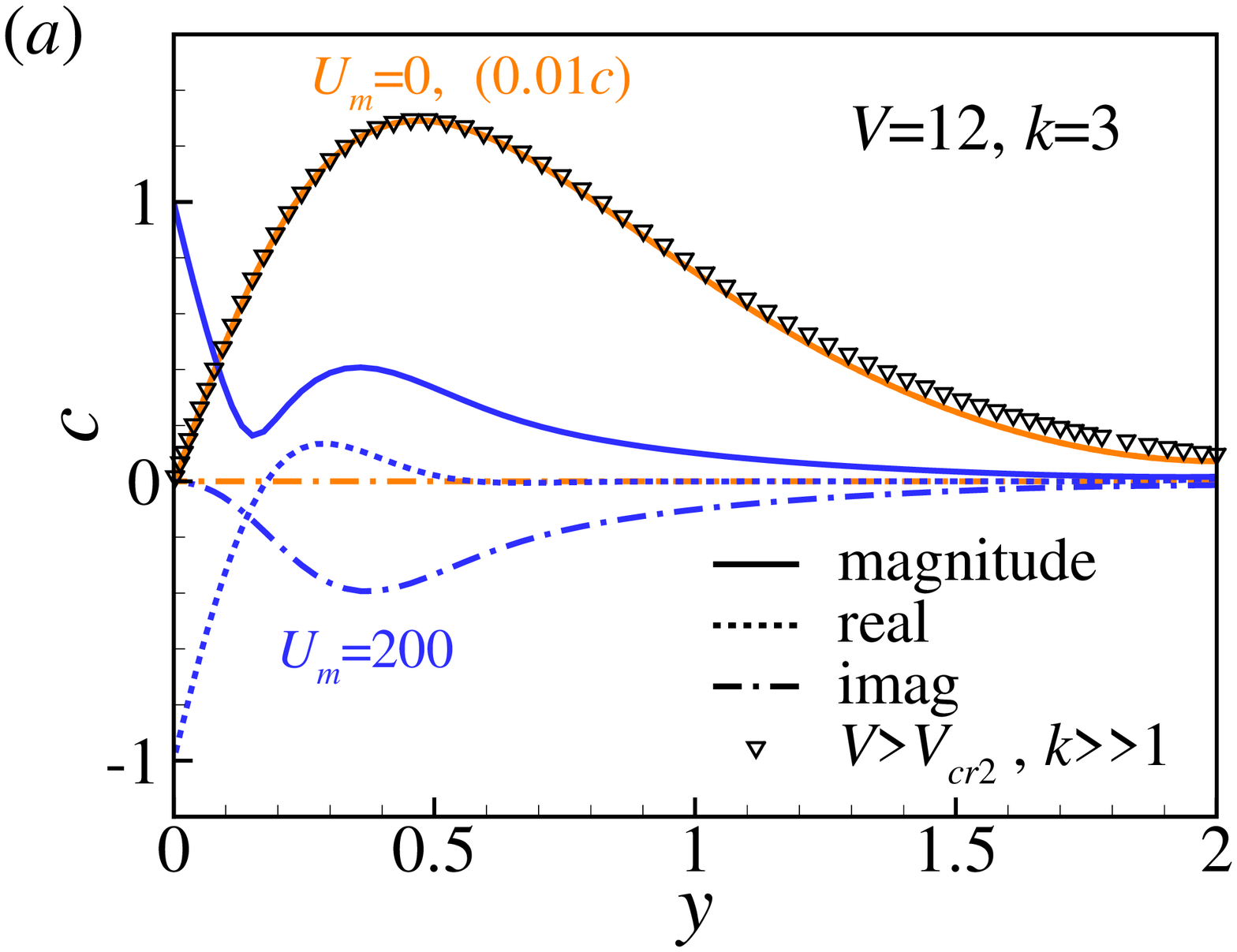}
\includegraphics[angle=0,scale=0.32]{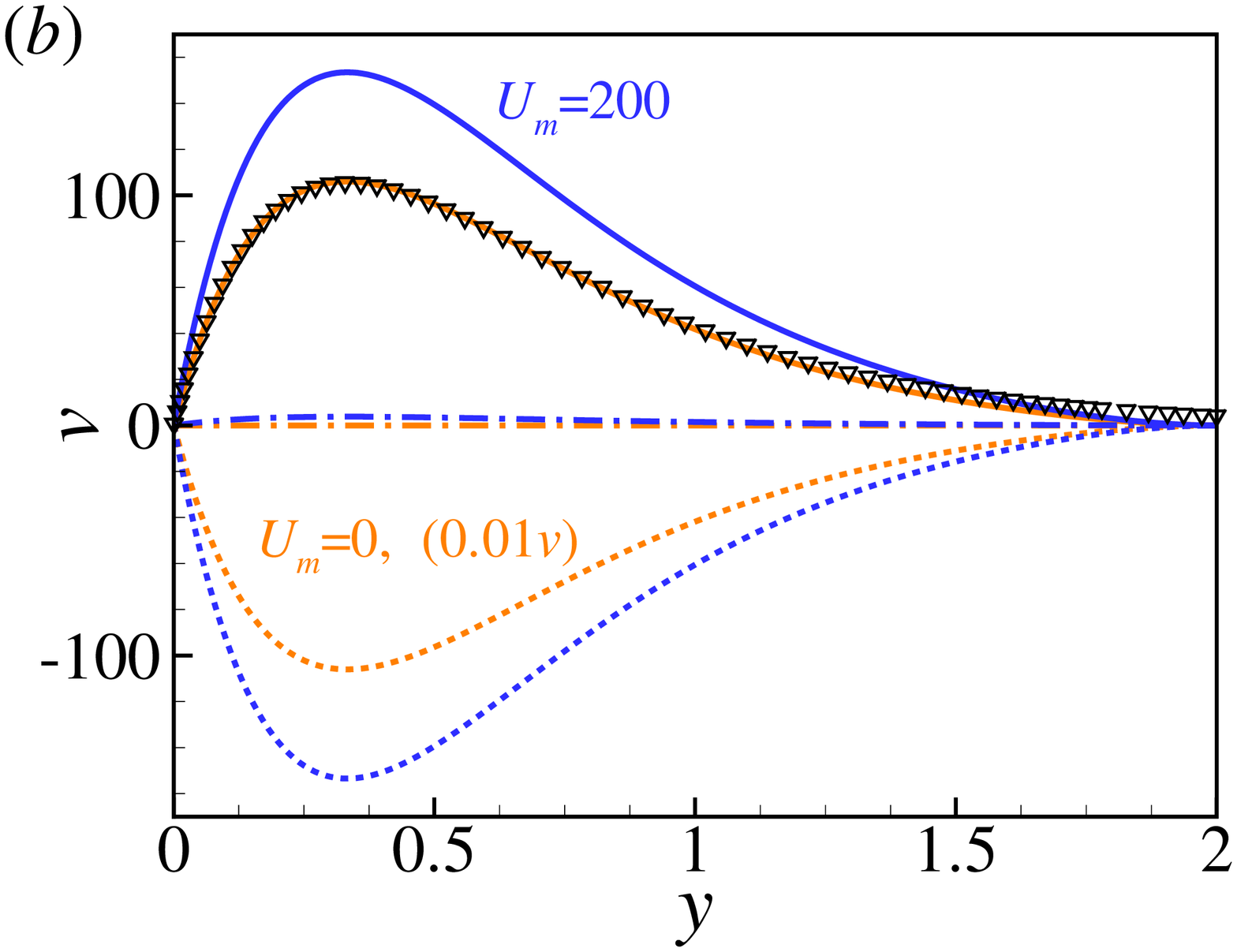}
\caption{Distribution of the perturbed ($a$) ion concentration and ($b$) normal velocity for the most unstable mode of the coupled instability at $U_m=0$ and 200, $V=12, k=3$. The results for $U_m=0$ are multiplied by $0.01$. Symbols show the high wavenumber asymptotic solutions (\ref{eq:largek2b}).}\label{fig:fig8}
\end{center}
\end{figure}

The imposed flow stabilizes the base state for the coupled instability problem mainly through mitigating the electroconvection. Figure \ref{fig:fig8} shows the eigenfunctions of the most unstable mode for $V=12$ and $k=3$ at $U_m=0$, for which electroconvective instability occurs, and $U_m=200$, for which electroconvection is suppressed by the imposed flow. The amplitude of the perturbed electrode surface is $h=1$.
In both cases, the local peak of the anode surface causes a downward flow and an enhanced ion concentration gradient which brings more ions to the peak and amplifies the instability. This result is consistent with previous experimental observations that the flow converges at the tips of the dendrites ~\cite[]{fleury1993coupling}. The cross flow greatly reduces the ion concentration gradient and the velocity of the downward flow, therefore reducing the growth rate of the morphological instability.

%%%%%%%%%%%%%%%%%%%%%%%%%%%%%
\section{Full analysis}\label{sec:full}
%%%%%%%%%%%%%%%%%%%%%%%%%%%%%

\subsection{Method}

\begin{figure}
\begin{center}
\includegraphics[angle=0,scale=0.35]{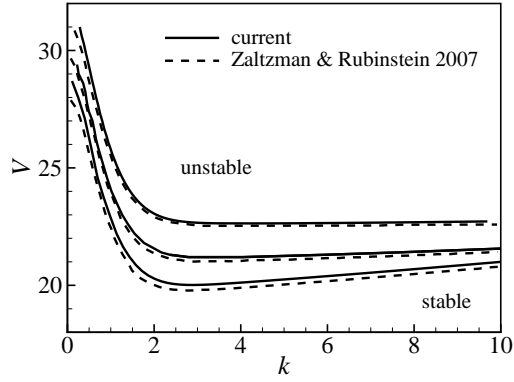}
\caption{The marginal stability curves for the purely electroconvective instabilities with different double layer thickness, $\mathrm{Pe}=0.5, D=1,$ and $U_m=0$. From bottom to top: $\delta=3\sqrt{2}\times10^{-4}$, $\sqrt{2}\times10^{-4}$, and $3\sqrt{2}\times10^{-5}$.  The solid lines are obtained with the ultraspherical method.  The dashed lines show the results from \cite{zaltzman2007electro}. }\label{fig:fig9}
\end{center}
\end{figure}

So far, we have considered the linear instability of the bulk region, assuming electroneutrality and using the second-kind electroosmotic slip velocity to replace the thin space charge layer. This simplification allows us to analytically derive the asymptotic solutions for small and large wavenumbers. The bulk analysis predicts that the imposed flow cannot completely suppress the electroconvective instability because the ion concentration disturbance caused by the slip velocity $u_s\sim k^2$ will always dominate the stabilizing effect by the imposed flow at large wavenumber.  However, the assumption of thin space charge layer in the bulk analysis is no longer valid at high wavenumber and it incorrectly predicts an infinite growth rate $\sigma\sim k^2$ as $k\to\infty$ for the electroconvective instability ~\cite[]{zaltzman2007electro}. To understand the effects of the imposed flow on modes at high wavenumbers, we now consider the linear instability of the full region, which includes the double layer and the space charge layer and we no longer assume electroneutrality. In the full analysis, the base state is derived by numerically solving Equ. (\ref{eq:fullbase}) and (\ref{eq:fbcbase}). The eigenmodes are then calculated by solving the perturbed equations
\begin{subequations}\label{eq:fullperturb}
\begin{equation}\label{eq:fullperturba}
\sigma c^+
+ik\mathrm{Pe}U_my(2-y)c^++\mathrm{Pe}vC^{+\prime}
=\frac{D+1}{2}\left(c^{+\prime\prime}
-k^2c^+-k^2C^+\phi+(C^+\phi'+c^+\Phi')'\right),
\end{equation}
\begin{equation}\label{eq:fullperturbb}
\sigma c^-
+ik\mathrm{Pe}U_my(2-y)c^-+\mathrm{Pe}vC^{-\prime}
=\frac{D+1}{2D}\left(c^{-\prime\prime}
-k^2c^-+k^2C^-\phi-(C^-\phi'+c^-\Phi')'\right),
\end{equation}
\begin{equation}\label{eq:fullperturbc}
2\delta^2(\phi''-k^2\phi)=c^--c^+,
\end{equation}
\begin{equation}\label{eq:fullperturbd}
v^{(4)}-2k^2v''+k^4v=k^2((\phi''-k^2\phi)\Phi'-\phi\Phi^{(3)}),
\end{equation}
\end{subequations}
with boundary conditions
\begin{subequations}
\begin{equation}\label{eq:fullperturbbc1}
(c^++C^{+\prime}h)|_{y=0}=0, \quad
c^+|_{y=2}=0,
\end{equation}
\begin{equation}
(c^{-\prime}-C^-\phi'-c^-\Phi')|_{y=0,2}=0,
\end{equation}
\begin{equation}
(\phi+\Phi'h)|_{y=0}=0, \quad
\phi|_{y=2}=0,
\end{equation}
\begin{equation}
v|_{y=0}=\frac{\sigma h}{\mathrm{Pe}}, \quad
v|_{y=2}=0, \quad
v'|_{y=0,2}=0,
\end{equation}
\begin{equation}\label{eq:fullperturbbc}
\frac{1+D}{2}(c^{+\prime}+C^+\phi'+c^+\Phi')|_{y=0}
=\frac{\sigma h}{v_m}.
\end{equation}
\end{subequations}
Both the base state and the perturbed equations are solved using the ultraspherical spectral method ~\cite[]{olver2013fast}. In contrast to the classical Chebyshev collocation method ~\cite[]{weideman2000matlab}, this method constructs the matrices in the coefficient space and uses banded operators to greatly reduce the condition number of the matrices. This allows inclusion of more Chebyshev coefficients to fully resolve the thin double layers. For the generalized eigenvalue problem, the ultraspherical spectral method is accurate up to around 6000 coefficients, while the classical Chebyshev collocation method becomes ill-conditioned with more than 100 collocation points. To validate the current method, we first consider the electroconvective instability of the full region without a flow.
Table \ref{tab:tab1} shows that the largest growth rates at different $k$ for the electroconvective instability derived by the ultraspherical spectral method and the shooting method for $\delta=10^{-3}, V=25$ and $V_m=0$ are nearly identical.  Figure \ref{fig:fig9} compares the marginal stability curves obtained with the ultraspherical method to those from a previous study that used a shooting method \cite{zaltzman2007electro}. Here $\mathrm{Pe}=0.5, D=1$ and $U_m=0$. The two results agree well with each other. The small differences are probably because in \cite{zaltzman2007electro} the equilibrium double layer near the top electrode surface is modeled with the first-kind electroosmotic slip velocity, while here it is fully resolved.

\begin{table}
\centering
\begin{tabular}{ c | c | c }
 $k$ & $\quad$Ultraspherical spectral method$\quad$ & Shooting method\\
 \hline
0.1 & -0.016096836493756 & -0.016096836696248 \\
1 & 1.279100815430633 & 1.279100816345562 \\
10 & 1.103806945236518e+02 & 1.103806944029065e+02 \\
100 & -9.996152401785079e+03 & $\quad$-9.996152401758340e+03$\quad$ \\
\end{tabular}
\caption{Comparison of the largest growth rates at different $k$ for the electroconvective instability derived using the ultraspherical spectral method and the shooting method, $\delta=10^{-3}, V=25, \mathrm{Pe}=0.35, D=2/3$ and $V_m=0$.}\label{tab:tab1}
\end{table}

\subsection{Electroconvection without imposed flow}

\begin{figure}
\begin{center}
\includegraphics[angle=0,scale=0.32]{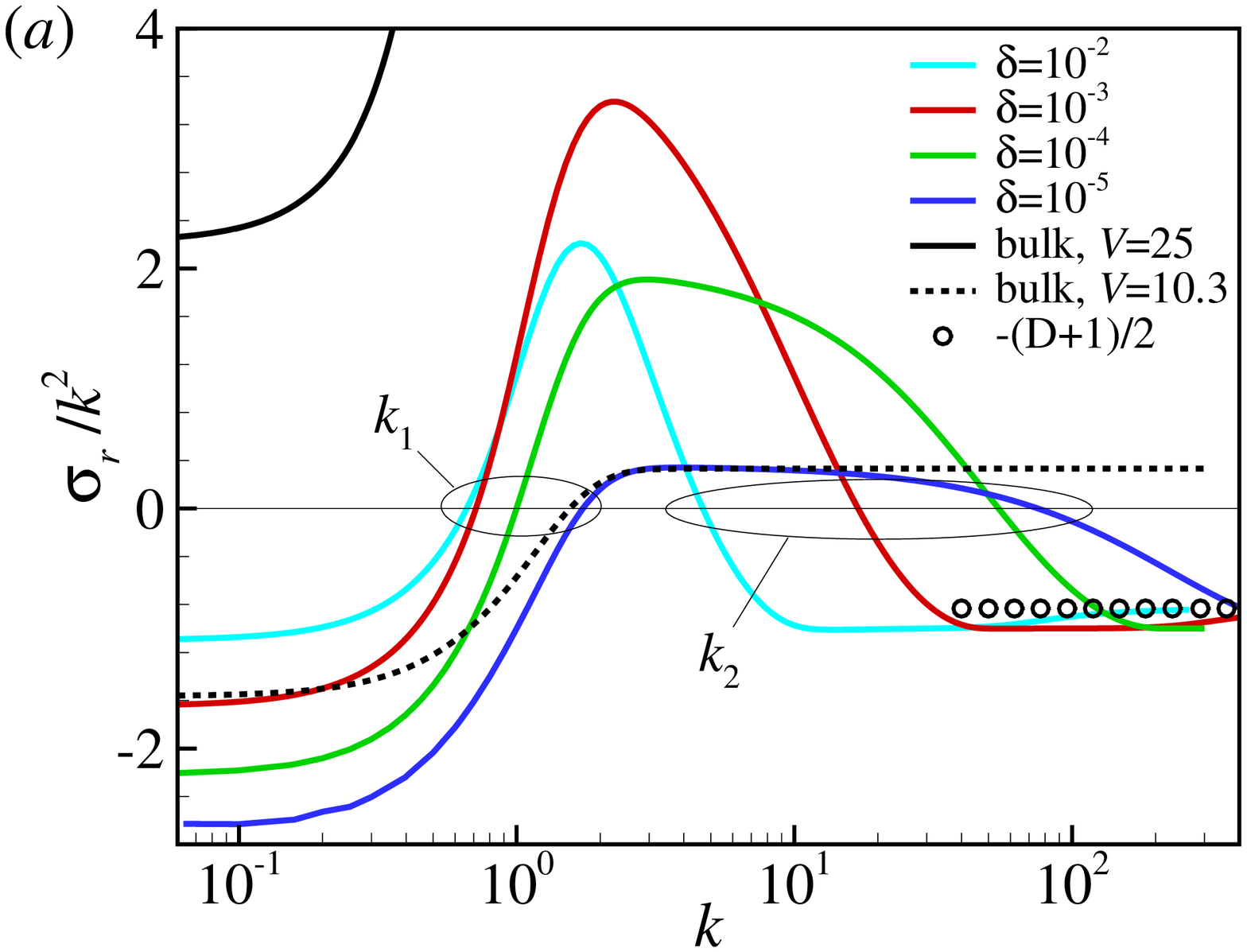}
\includegraphics[angle=0,scale=0.32]{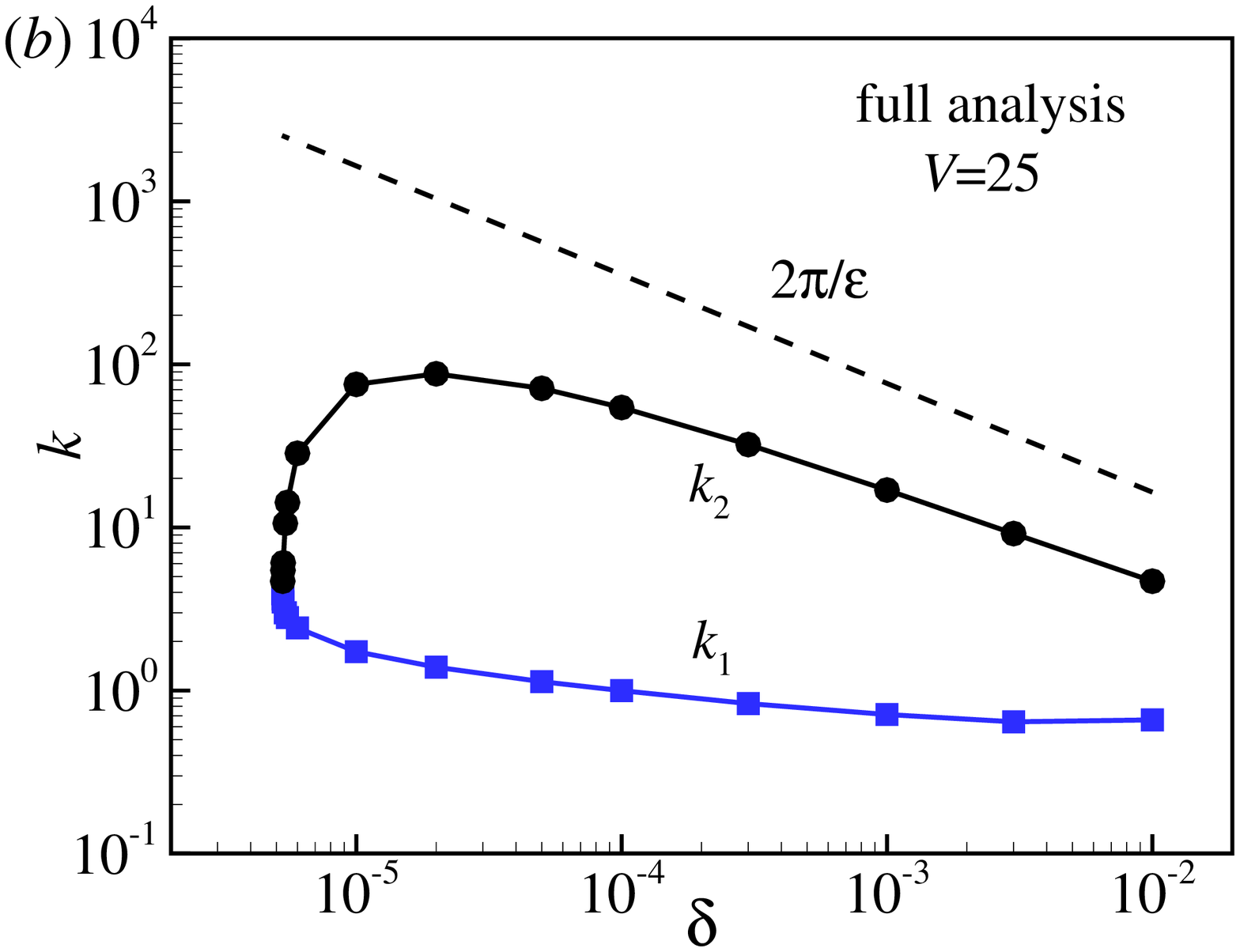}
\caption{($a$) The dependence of the growth rate $\sigma_r/k^2$ on the wavenumber $k$ for the purely electroconvective instability at $U_m=0, \mathrm{Pe}=0.35$ and $D=2/3$. Colored lines: full analysis at $V=25$ for different double layer thickness. Black lines: bulk analysis at different voltages. The solid black line ($V=25$) reaches $\sigma_r/k^2=16.89$ at large $k$, the dotted line ($V=10.3$) has better agreement with the full analysis at $\delta=10^{-5}$ for $1<k<10$. $k_1$ and $k_2$ represent the two wavenumbers at which the growth rate becomes zero. Open circles show the asymptotic solution $\sigma_r=-(D+1)k^2/2$ as $k\to\infty$. ($b$) The dependence of $k_1$ and $k_2$ on $\delta$ for the full analysis at $V=25$. The dashed line shows the corresponding wavenumber of the space charge layer, $\delta_s=(9\delta^2V^2/8)^{1/3}$.}\label{fig:fig10}
\end{center}
\end{figure}

Before considering the effects of the imposed flow, we first compare the electroconvective instabilities derived from the bulk and full analysis. Fig. \ref{fig:fig10}($a$) shows the largest growth rate of the purely electroconvective instability with different $\delta$ at $U_m=0$. At $V=25$, the full analysis predicts stable modes at small ($k<k_1$) and large ($k>k_2$) wavenumbers and unstable modes in between. At $V=25$, the bulk analysis predicts unstable modes at all wavenumbers since the voltage is above the critical voltages $V_{cr1}=14.4$ for $k\ll1$ and $V_{cr2}=9.6$ for $k\gg1$. Its predicted growth rates $\sigma_r=2.22k^2$ for $k\ll1$ and $\sigma_r=16.89k^2$ for $k\gg1$ are much larger than those from the full analysis at the same voltage. One reason for this result is that the bulk analysis neglects the $\mathrm{O}(\ln\delta)$ potential drop inside the double layer and space charge layer ~\cite[]{rubinstein2001electro} and therefore overestimates the slip velocity. Reducing the voltage to smaller values decreases the growth rate for the bulk analysis. For example, adjusting the voltage to $V=10.3$ brings the bulk analysis for $1<k<10$ closer to the full analysis for $\delta=10^{-5}$ as shown in figure \ref{fig:fig13}. Later, we will see that these two analyses also have similar eigenfunctions.
The bulk analysis at $V=10.3$ still quickly deviates from the full analysis at large $k$ and predicts an infinite growth rate as $k\to\infty$.
The deviation occurs at a wavenumber which is smaller than the inverse of the space charge layer $\varepsilon\sim(\delta V)^{2/3}$ ~\cite[]{rubinstein2005electroconvective}, showing that the linear instability of the mode is strongly influenced by the space charge layer. For $k<k_1$ and $k>k_2$, the full analysis predicts a stable mode with $\sigma_r\sim-k^2$, indicating that the mode is stabilized by diffusion of the ions inside the space charge layer. As $k\to\infty$, $\sigma_r\sim-(D+1)k^2/2$.
For an intermediate range of wavenumbers, $k_1<k<k_2$, the full analysis predicts an unstable mode whose growth rate scales as $\sigma_r\sim k^2$ due to the scaling of the electroosmotic slip velocity. The result is more evident at smaller $\delta$ and larger $V$, where the perturbation is unstable over a larger range of wavenumbers.
To better understand the transitions between stable and unstable modes, we plot $k_1$ and $k_2$ as functions of $\delta$ at $V=25$ in Fig. \ref{fig:fig10}($b$). The result is composed of two regions.
At large $\delta$, the applied voltage $V=25$ is well above the critical voltage for the onset of the electroconvective instability. $k_1\sim$O(1) and it slightly increases with decreasing $\delta$, suggesting that the stabilization of the modes at small wavenumbers is related to the gap between the two electrode surfaces. $k_2$ roughly scales as $\delta^{-2/3}$, meaning that the modes are stabilized due to the space charge layer whose thickness is $\delta_s=(9\delta^2V^2/8)^{1/3}$ ~\cite[]{chazalviel1990electrochemical}.  This stabilization cannot be captured by the bulk analysis.
At smaller $\delta$, when the applied voltage $V=25$ is close to the critical voltage, $k_1$ and $k_2$ change more rapidly with decreasing $\delta$ and eventually $k_1=k_2$ reaching the critical wavenumber for neutral instability. At small $\delta$, $k_1$ and $k_2$ do not follow a simple scaling law with the gap thickness or the space charge layer thickness, probably because the potential drop across the double layer and space charge layer has a large effect.

\begin{figure}
\begin{center}
\includegraphics[angle=0,scale=0.32]{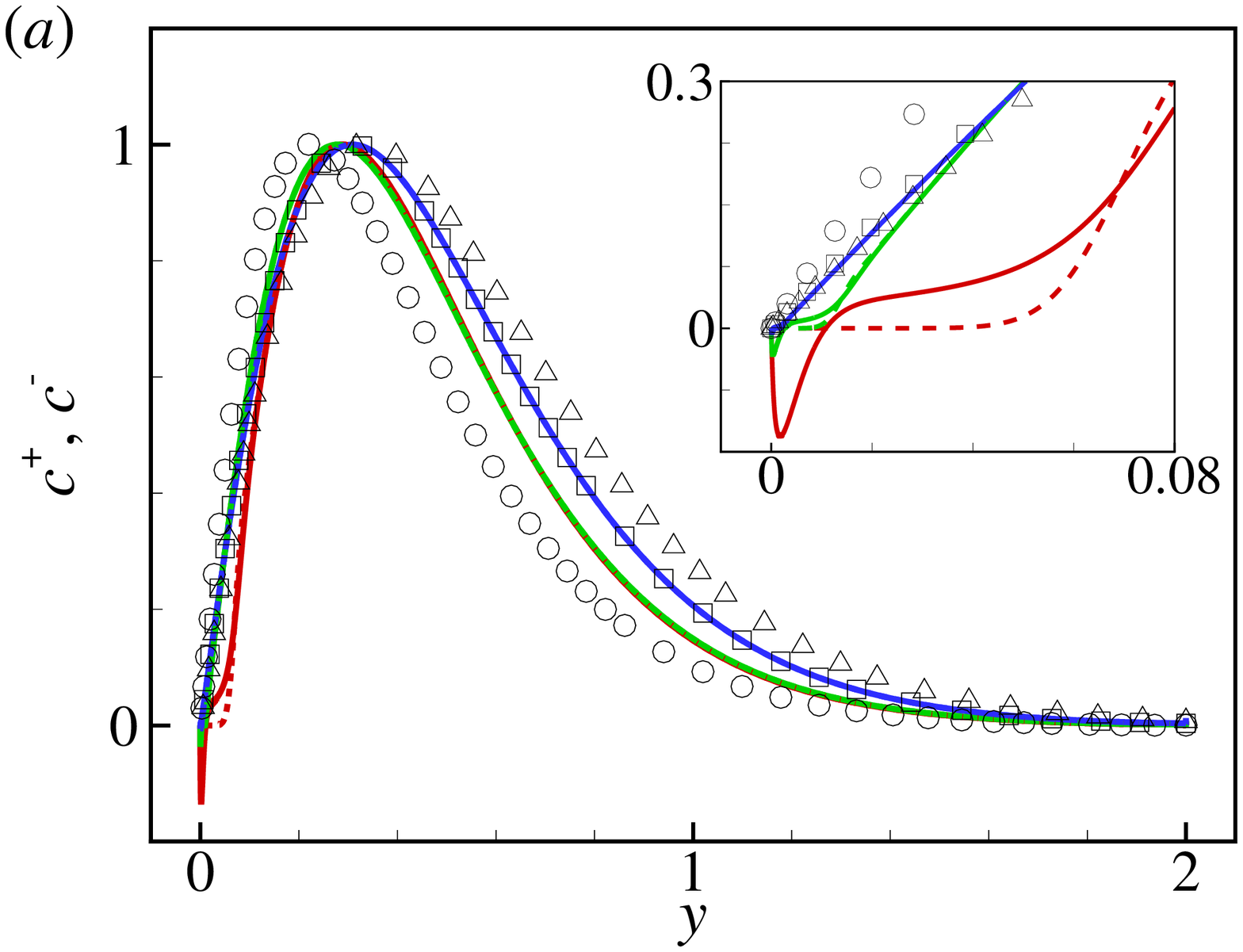}
\includegraphics[angle=0,scale=0.32]{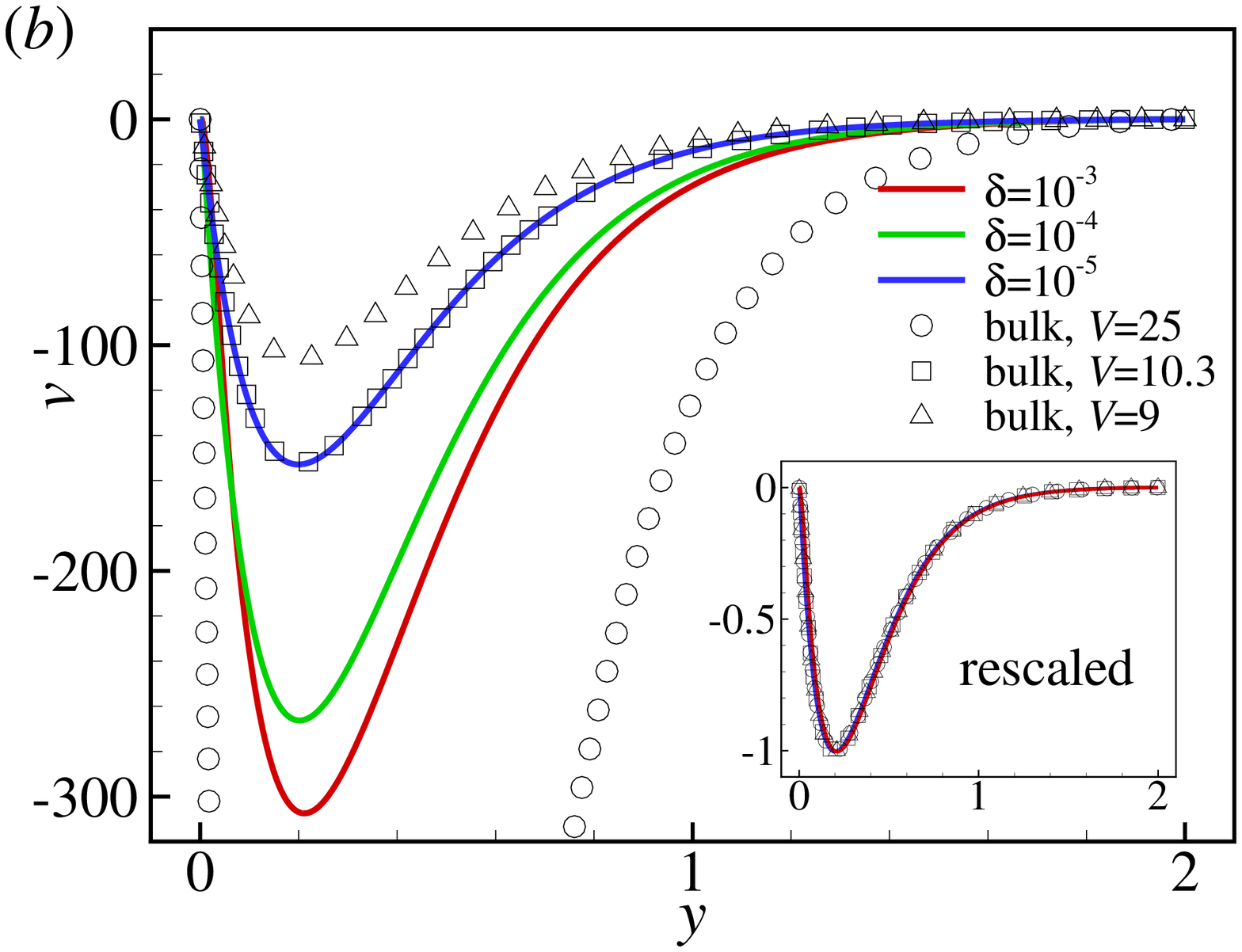}
\caption{The perturbed ($a$) ion concentration (solid: cation, dashed: anion) and ($b$) velocity for the electroconvective instability at $U_m=0$ and $k=5$. Lines: full analysis at $V=25$. Symbols: bulk analysis at different voltages.}\label{fig:fig11}
\end{center}
\end{figure}

The similarities between the bulk and full analyses for $k<k_2$ are further illustrated through their eigenfunctions in fig. \ref{fig:fig11}. Here, we compare the profiles of ion concentrations $c^+, c^-$ and normal velocity $v$ at $k=5$. The eigenfunctions are normalized such that the ion concentrations have the same peak value of $c^+_{max}=1$.
The overall distribution of the perturbed ion concentration is not sensitive to the specific values of $\delta$ or $V$ in fig. \ref{fig:fig11}($a$). However, the ion concentration distribution and its gradient near the space charge layer, which determines the slip velocity and the electroconvective instability, is strongly dependent on $\delta$ and $V$.
In the full analysis, the variation of the cation ion concentration in the space layer at $y=0$ decreases with decreasing $\delta$. Since the velocity field is driven by $(c^--c^+)/\delta^2$, the effects of the space charge layer on the ion concentration is not negligible even at small $\delta$. The magnitude of the slip velocity and the instability is determined by the slope of the ion concentration outside the space charge layer. The bulk analysis at $V=25$ overestimates its gradient near $y=0$ and therefore leads to a much larger velocity that causes a larger growth rate. By reducing the voltage to $V=10.3$, the ion concentration and velocity profiles agree well with the full analysis results for $\delta=10^{-5}$ leading to a closer agreement of the growth rate as well. All the velocities follow the same relation $v\sim ye^{-ky}$, meaning that the electroconvective instability is driven by osmotic slip velocities of the same form but with different magnitudes.

\begin{figure}
\begin{center}
\includegraphics[angle=0,scale=0.4]{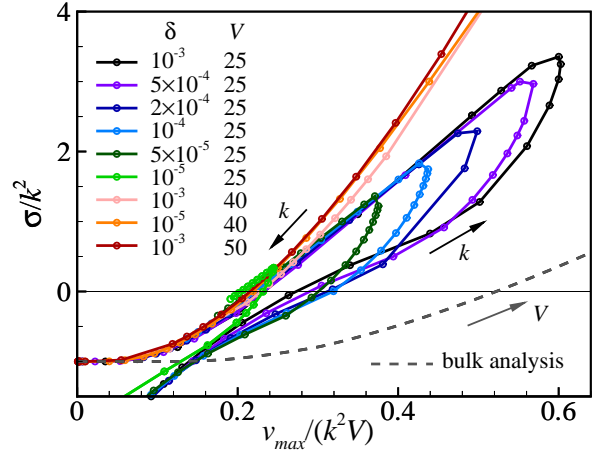}
\caption{The dependence of $\sigma_r/k^2$ on $v_{max}/(k^2V)$. $v_{max}$ is the maximum normal velocity when the perturbed ion concentration has the peak value $c^+_{max}=1$. Symbol-lines show the full analysis results, each individual symbol corresponds to a specific wavenumber $k$. Increasing $k$, the transitions from stable to unstable then to stable modes again correspond to the wavenumbers $k_1$ and $k_2$, respectively. Dotted line shows the bulk analysis results at $k=50$ with the voltage $V$ increasing from left to right. The bulk analysis is derived from equations (\ref{eq:largek1v}), (\ref{eq:largek1a}) and (\ref{eq:largek1c}).}\label{fig:fig12}
\end{center}
\end{figure}

\begin{figure}
\begin{center}
\includegraphics[angle=0,scale=0.32]{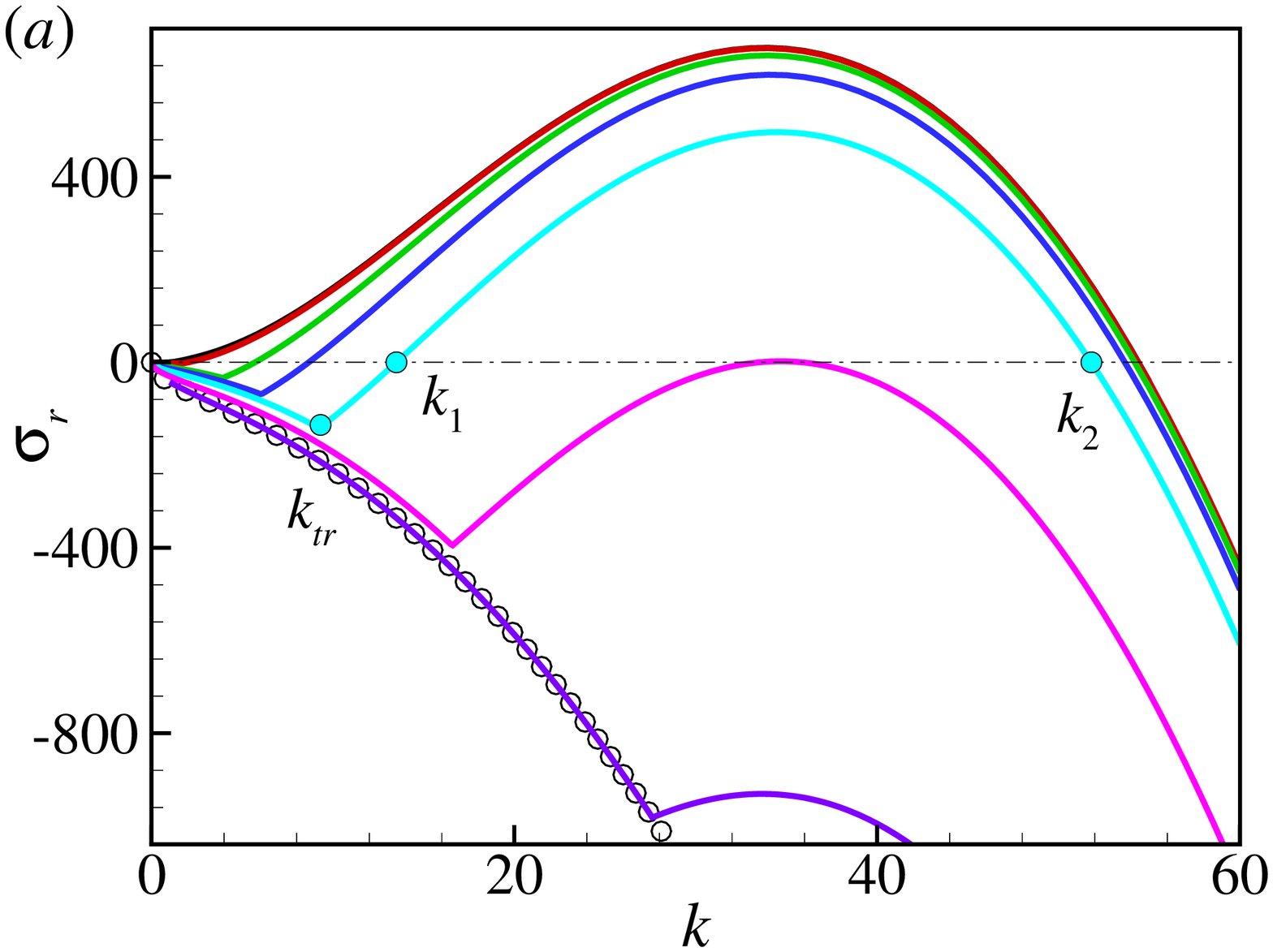}
\includegraphics[angle=0,scale=0.32]{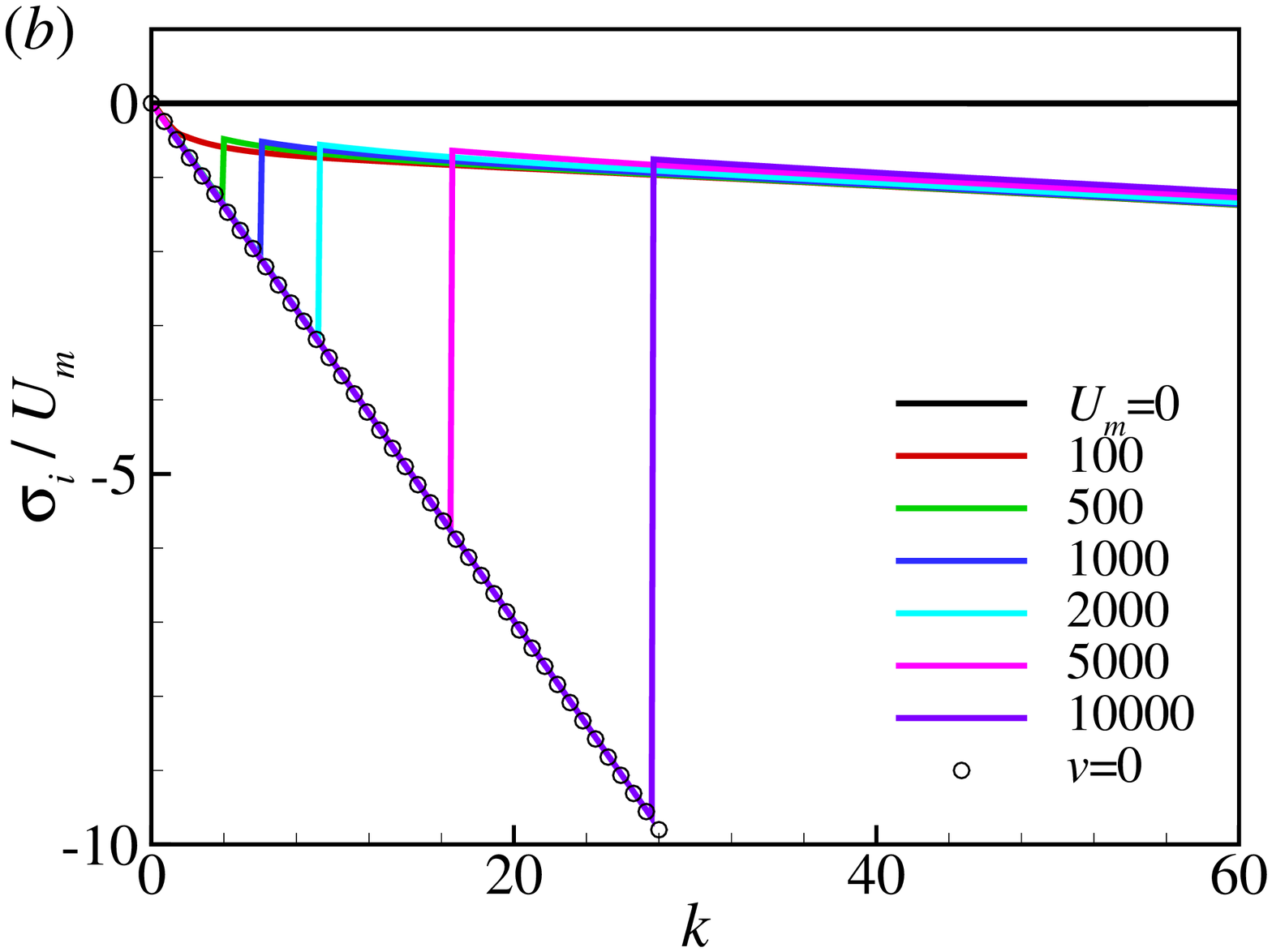}
\caption{The ($a$) real and ($b$) imaginary parts of the maximum growth rate of the purely electroconvective instability at different $U_m$ for $\delta=10^{-4}$ and $V=25$. Symbols show the analytical solution for $v=0$ given in equ. (\ref{eq:vequ0}). $k_{tr}$ is the transition wavenumber between the centerline and wall modes. $k_1$ and $k_2$ are the wavenumbers for zero growth rate.}\label{fig:fig13}
\end{center}
\end{figure}

For the unstable mode, the bulk analysis can be made quantitatively comparable with the full analysis by adjusting the voltage. However, the bulk analysis cannot predict the transition from an unstable to a stable mode with increasing $k$. To better understand this transition at $k=k_2$, we plot the rescaled growth rate $\sigma_r/k^2$ as a function of $v_{max}/(k^2V)$ in fig. \ref{fig:fig12}. Since the eigenfunctions can be multiplied by any constant without affecting the result, we normalized the eigenfunctions such that the ion concentration has the peak value $c^+_{max}=1$. The symbol-lines show the results of the full analysis and each individual point represents a specific wavenumber $k$.
The growth rate strongly depends on the magnitude of the rescaled velocity $v_{max}/(k^2V)$. The transition from stable to unstable mode at $k_1$ depends on $\delta$ and $V$. For $\delta\geq5\times10^{-5}, V=25$, it occurs at $v_{max}/(k^2V)\simeq0.3\pm0.02$, for $\delta=10^{-5}, V=25$ which is closer to the neutral stability, it occurs at $v_{max}/(k^2V)\simeq0.23$, and at higher voltages, all small wavenumber perturbations are unstable and $k_1$ does not exists. In comparison, the transition from unstable to stable mode at $k_2$ always occurs at $v_{max}/(k^2V)\simeq0.21\pm0.01$. This result shows that the electroconvective instability is suppressed if the normal velocity is not strong enough to overcome the ion diffusion and sustain the ion flux from low to high concentration regions.
The bulk analysis only shows the transition at $k_1$. It overestimates the transition velocity $v_{max}/(k^2V)=(\sqrt{5}-2)e^{(\sqrt{5}-1)/2}/\sqrt{2\mathrm{Pe}}=0.52348$, where the numerical factor is calculated from the bulk analysis prediction for $k\gg1$. This is because it underpredicts the voltage for the onset of the electroconvective instability \cite[]{zaltzman2007electro}.

\subsection{Effect of imposed flow on electroconvection}

We now consider the effects of the cross-flow on the electroconvective instability. Fig. \ref{fig:fig13} shows the complex eigenvalue $\sigma$ as a function of $k$ at different $U_m$, $\delta=10^{-4}$ and $V=25$. Below the transition wavenumber $k_{tr}$, the perturbation is stable and it propagates downstream with the velocity at the channel centerline $PeU_m$. The eigenvalue of the mode follows the same asymptotic solution (\ref{eq:vequ0}) in the bulk analysis. Above $k_{tr}$, the electroconvective instability is determined by the wall mode, which has a local maximum growth rate at $k\simeq34$ and the wave speed roughly follows $u_s\sim1/k$. The predictions of the bulk (see figure (\ref{fig:fig4})) and full analyses for the complex growth rate at relatively small wavenumbers are very similar to one another.
Increasing $U_m$ delays the transition from the center mode to the wall mode and reduces the peak value of the growth rate. At a fixed voltage, we also find the transition wavnumber scales as $k_{tr}\sim U_m^{1/2}$ as in the bulk analysis.
In the full analysis, full suppression of the electroconvective instability is possible in a strong enough flow. This result is qualitatively different from the bulk analysis.

\begin{figure}
\begin{center}
\includegraphics[angle=0,scale=0.32]{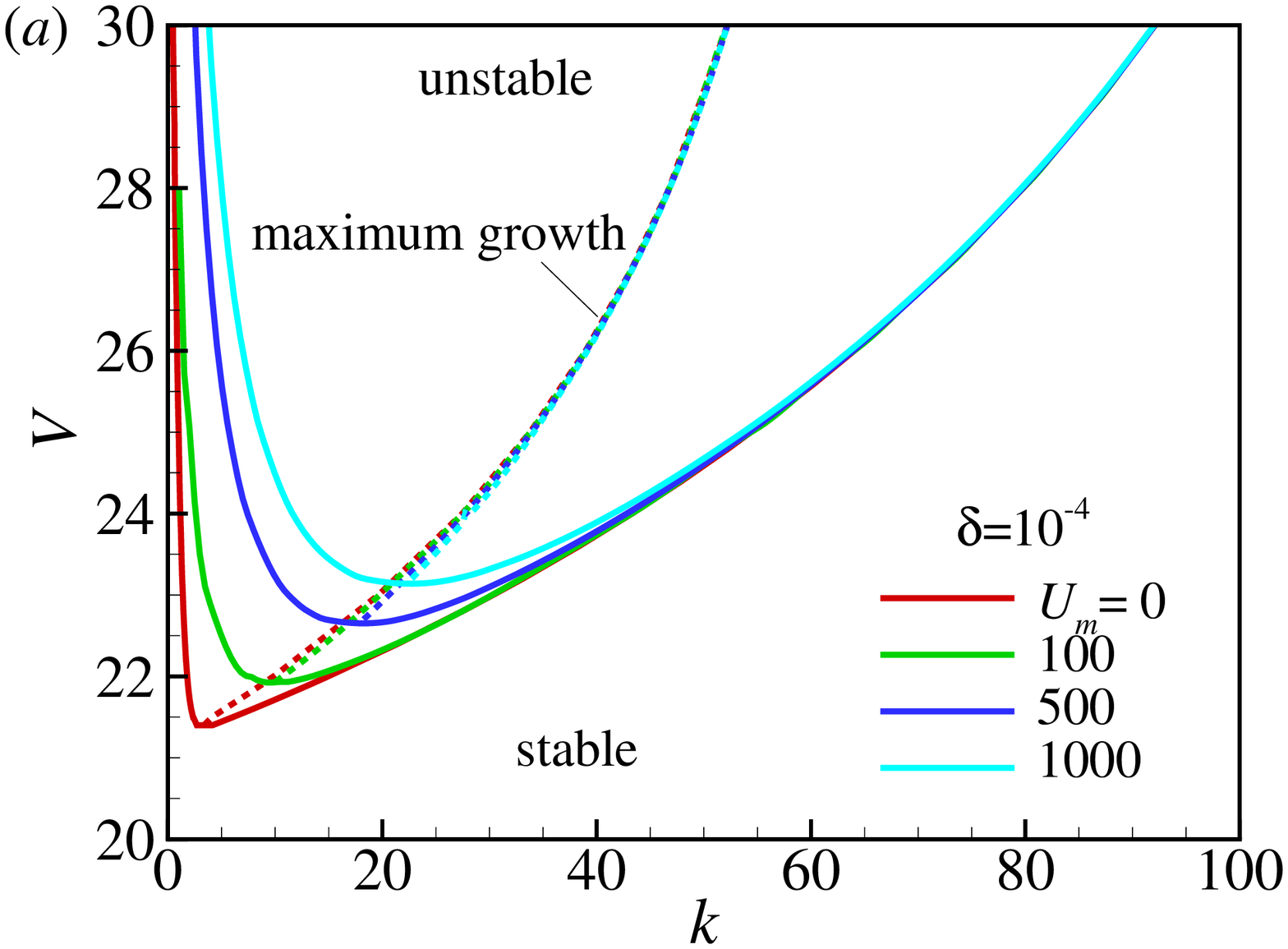}
\includegraphics[angle=0,scale=0.32]{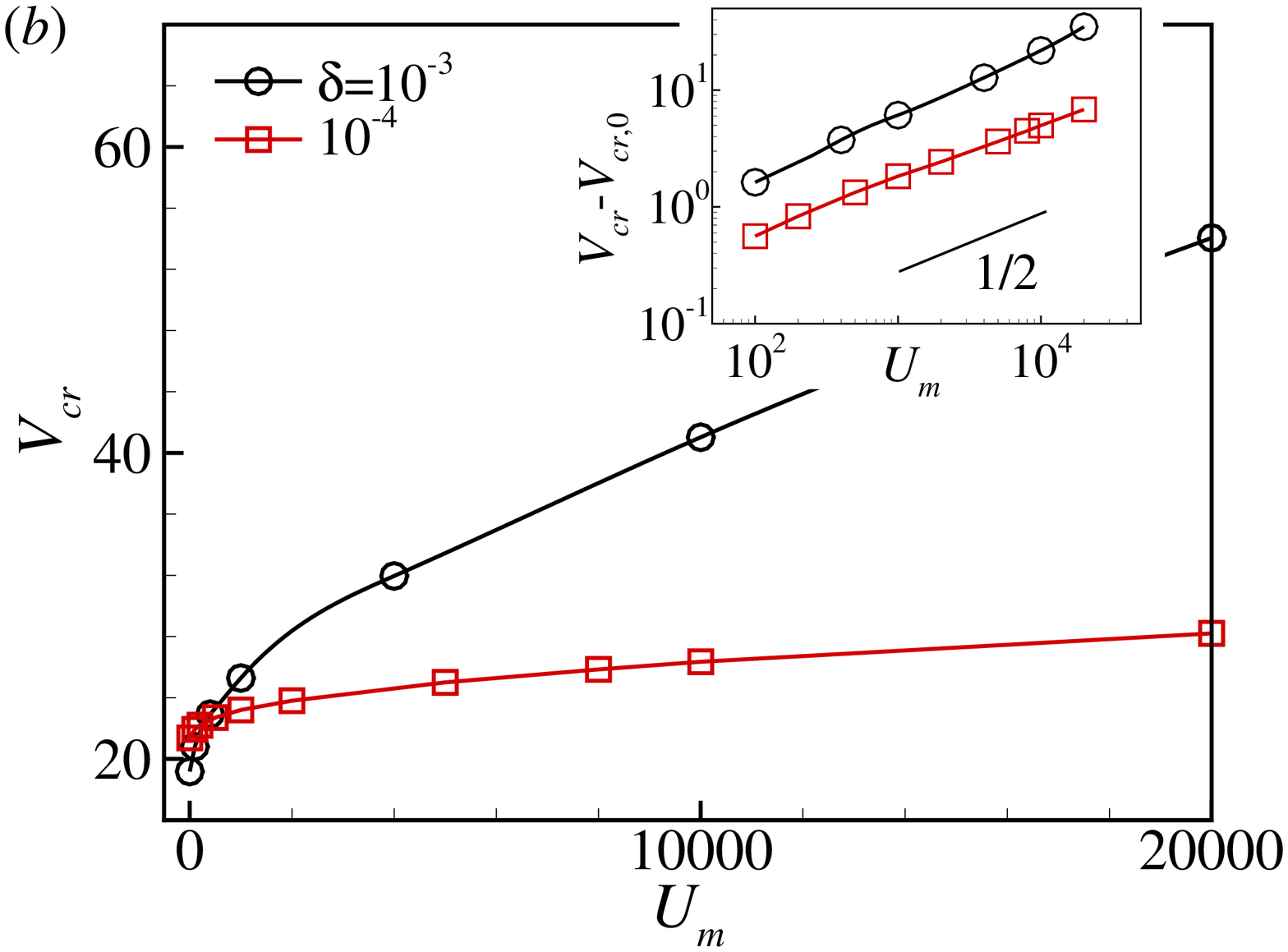}
\caption{($a$) Marginal stability curves for the purely electroconvective instability at different $U_m$, $\delta=10^{-4}$. ($b$) Dependence of the critical voltage $V_{cr}$ on the velocity of the cross-flow $U_m$ at different double layer thicknesses.}\label{fig:fig14}
\end{center}
\end{figure}

Fig. \ref{fig:fig14}($a$) shows the marginal stability curves at different imposed velocities $U_m$ for $\delta=10^{-4}$. The imposed flow stabilizes the modes at small wavenumbers, while it does not affect the transition at large wavenumbers or the wavenumber of the most unstable mode. In fig. \ref{fig:fig14}($b$), the critical voltage $V_{cr}$ for the onset of the electroconvective instability increases with increasing $U_m$. At the same velocity, the increase of the critical voltage is more prominent when the double layer thickness $\delta$ is larger, meaning that suppression of electroconvection by a cross-flow is more effective in an electrolyte with low salt concentration. This is because at the same voltage, the wavenumber of the most unstable mode decreases with increasing $\delta$ (see figure \ref{fig:fig10}), and the imposed flow is more effective in suppressing the instabilities at small wavenumbers. The voltage difference roughly scales as $V_{cr}-V_{cr, 0}\sim U_m^{1/2}$, where $V_{cr, 0}$ is the critical voltage at $U_m=0$.

\begin{figure}
\begin{center}
\includegraphics[angle=-90,scale=0.32]{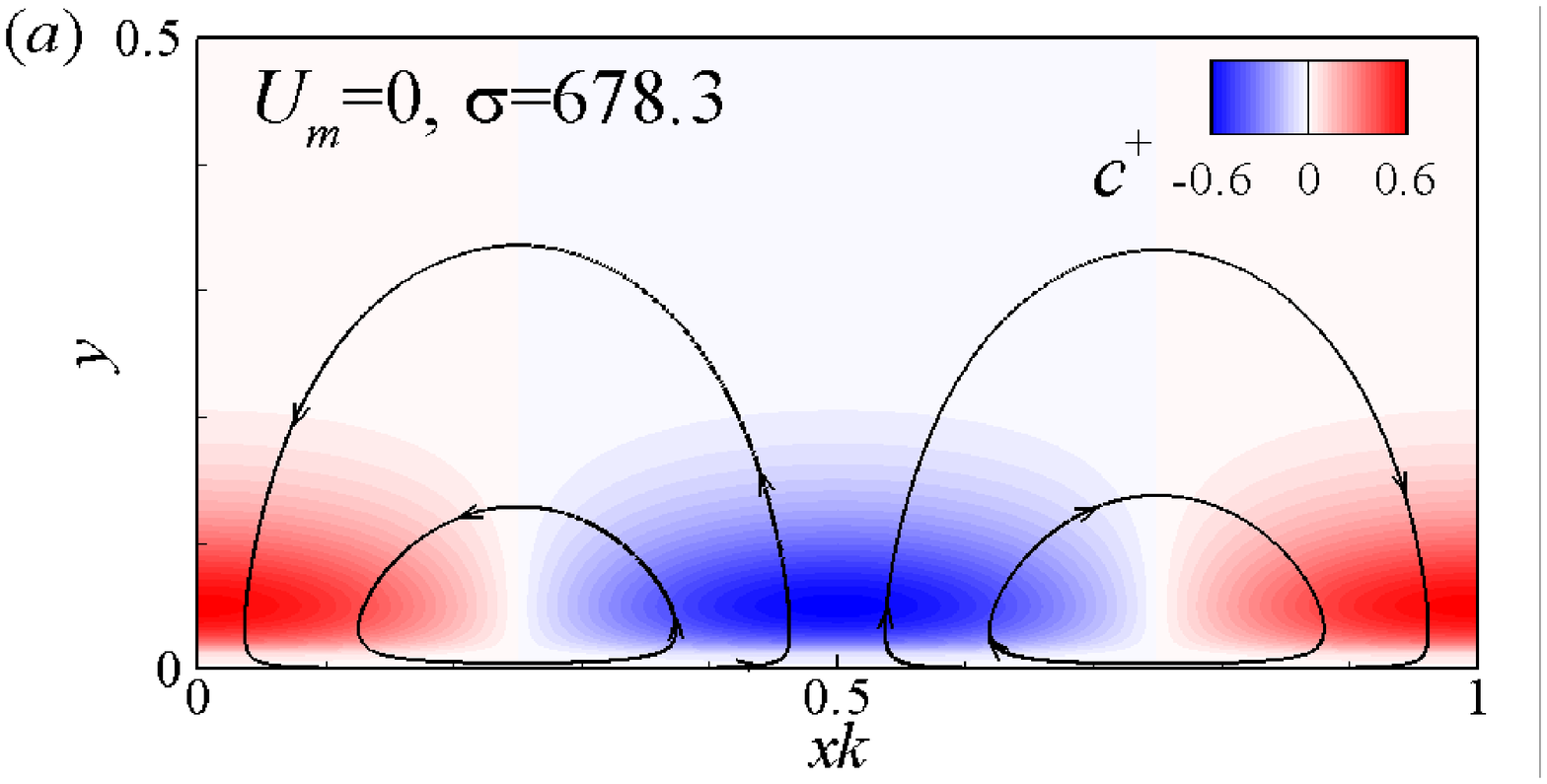}
\includegraphics[angle=-90,scale=0.32]{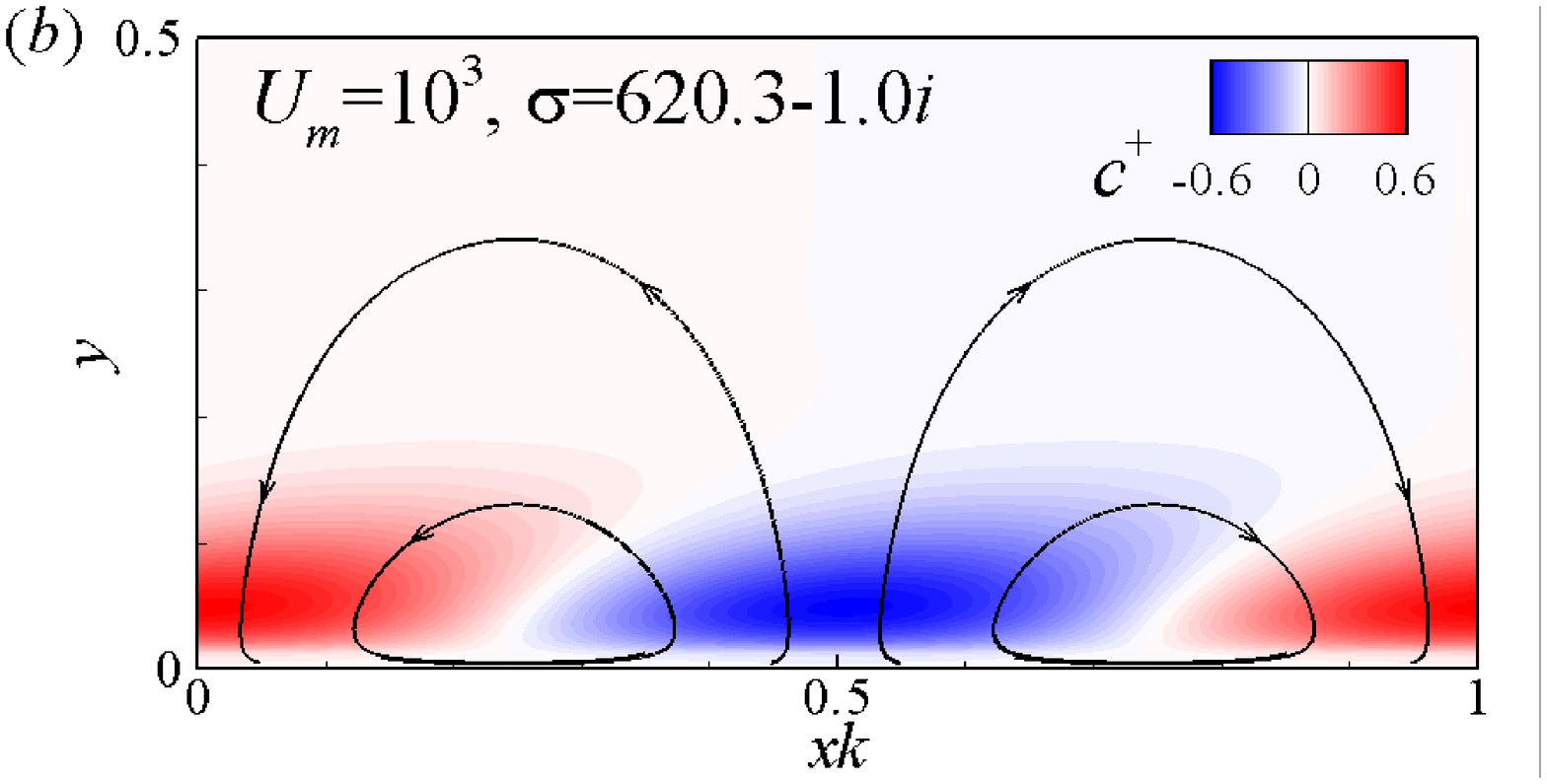}
\includegraphics[angle=-90,scale=0.32]{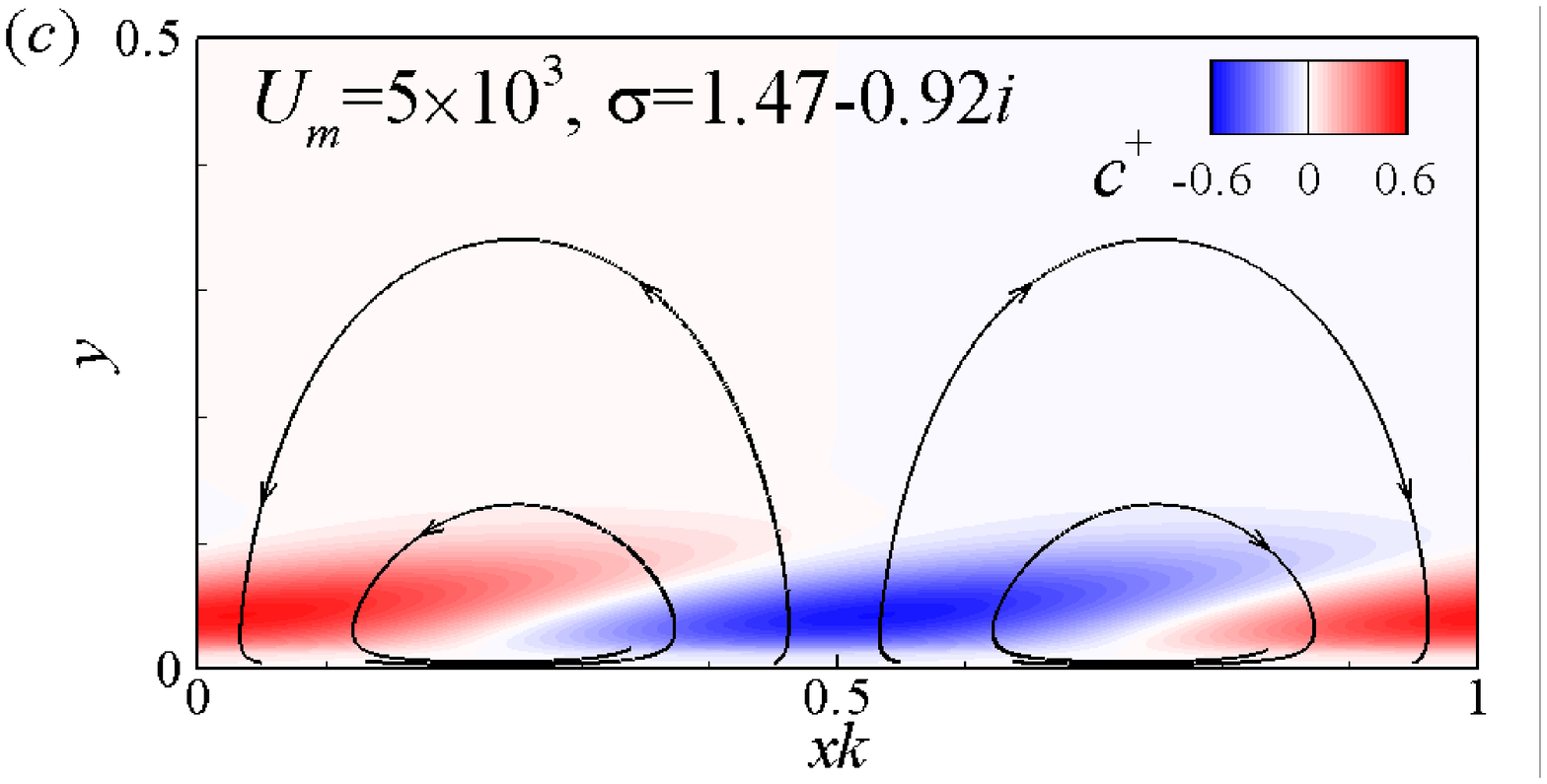}
\includegraphics[angle=-90,scale=0.32]{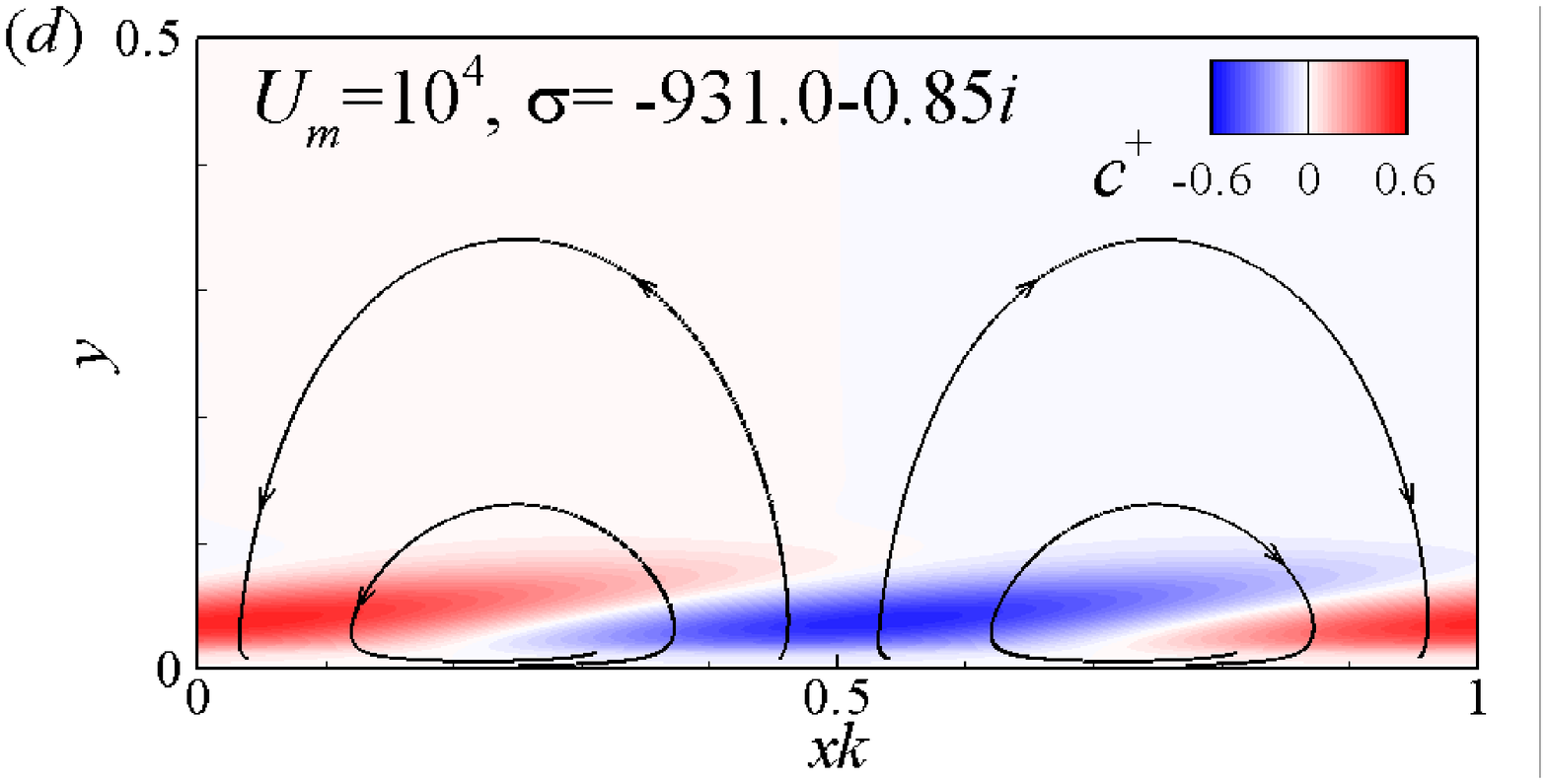}
\caption{Cation concentration distribution and the streamlines of the perturbation fluid velocity in the purely electroconvective instability at different imposed velocities, $\delta=10^{-4}, V=25$ and $k=34$.}\label{fig:fig15}
\end{center}
\end{figure}

The mechanism by which the cross-flow stabilizes the base state and suppresses electroconvection can be understood from the eigenfunctions in fig. \ref{fig:fig15}. Here, the plots are normalized such that all the results have the same maximum normal velocity. At $U_m=0$, both ion concentration and fluid velocity have real eigenfunctions, corresponding to stationary vortices which bring more ions from low to high concentration regions and therefore cause the instability when the convective flux is stronger than the stabilizing effect of ion diffusion.
The imposed flow has a small influence on the streamlines of the disturbance flow and its main effect is on the concentration field. The shear flow near the wall deforms the ion concentration and generates an inclined shielded region which hinders the ion flux from low to high concentration regions. With increasing velocity, the stretching of the perturbed ion concentration field becomes stronger and it eventually stabilizes the perturbation. We also notice that the magnitude of the perturbed ion concentration field increases with increasing $U_m$, which is consistent with the bulk analysis shown in figure (\ref{fig:fig5}). This result shows that the strength of the normal flow has a minor effect on the electroconvective instability in an imposed flow, in contrast to the cases without the flow.

\subsection{ Coupled electroconvective and morphological instability}

\begin{figure}
\begin{center}
\includegraphics[angle=0,scale=0.32]{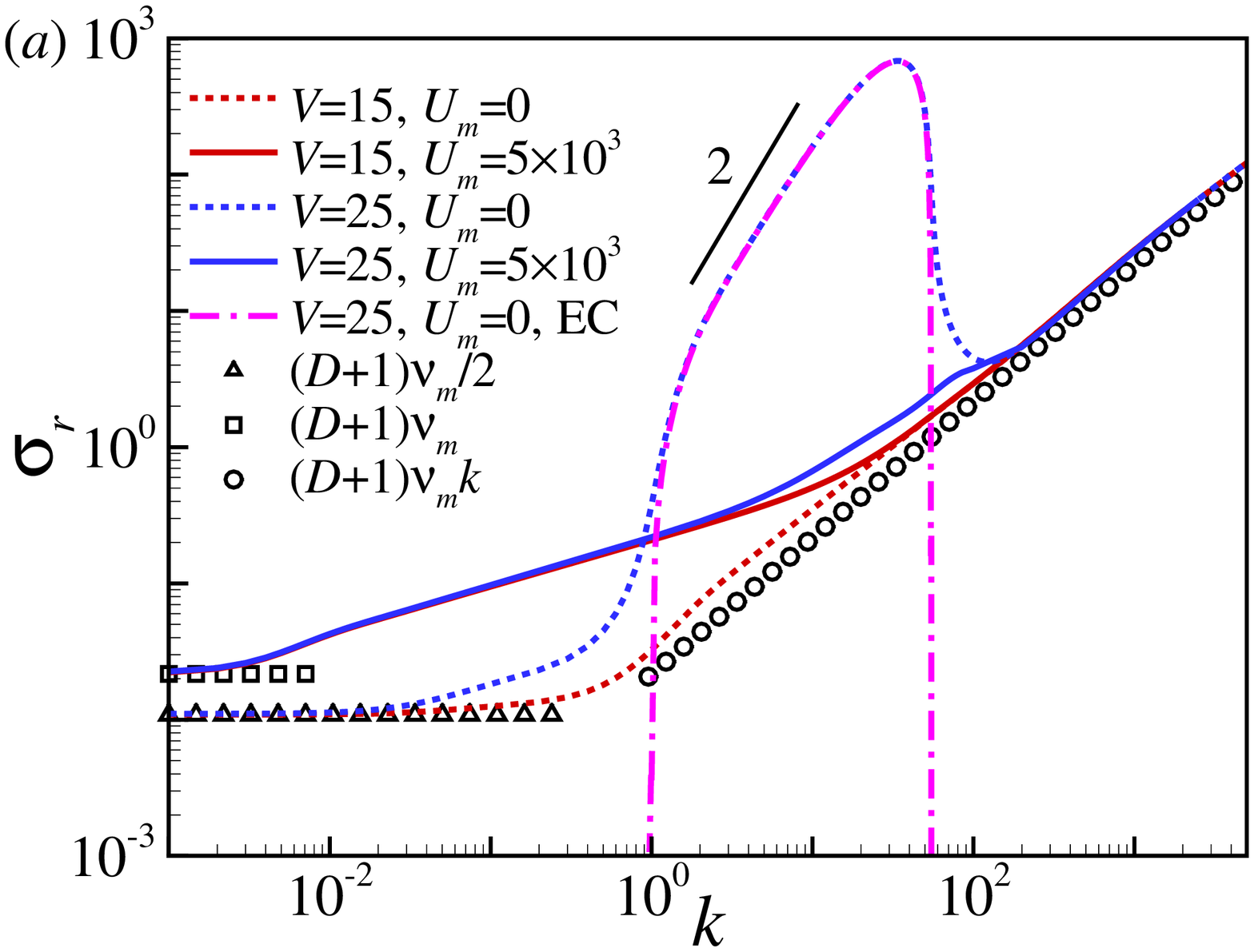}
\includegraphics[angle=0,scale=0.32]{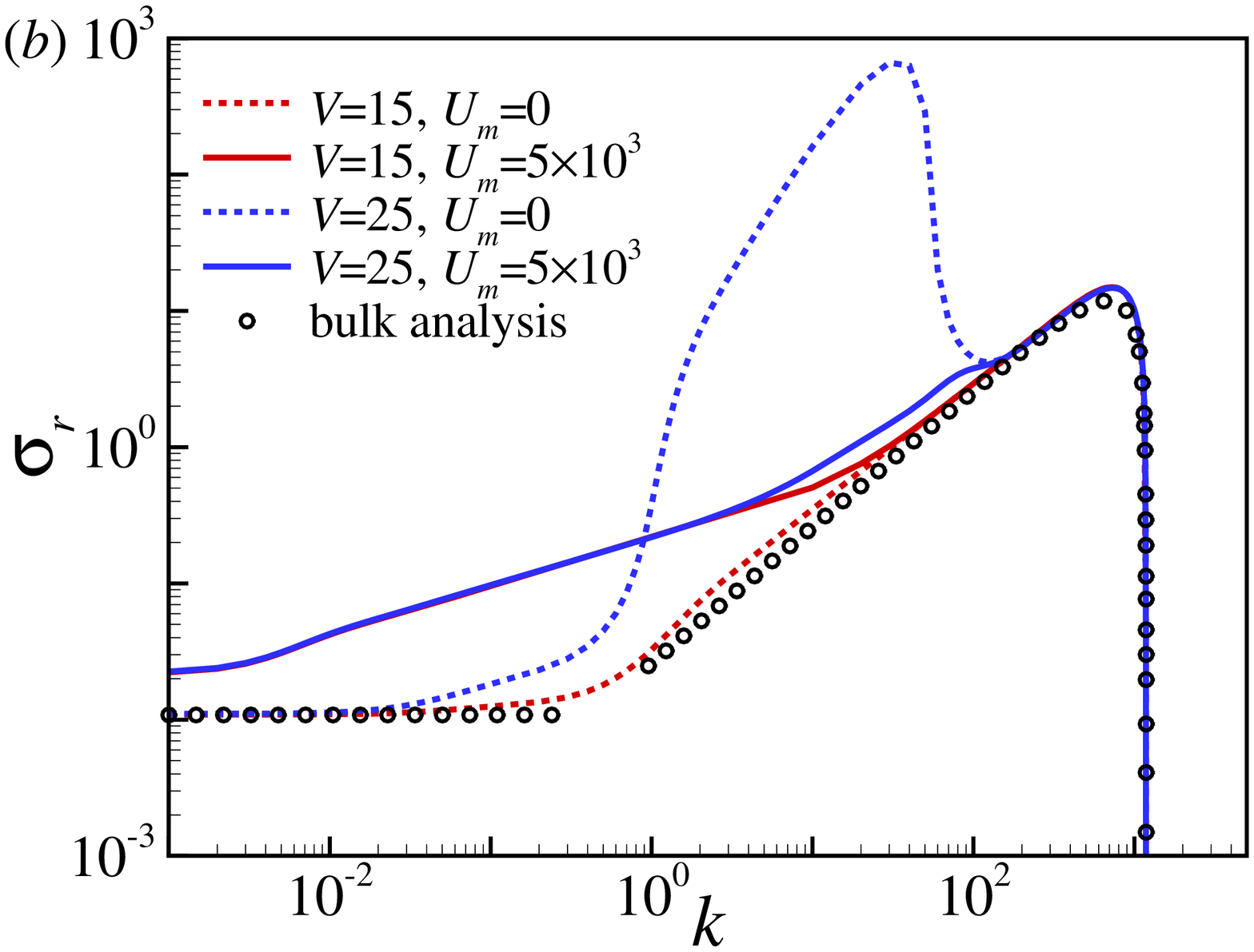}
\caption{($a$) Maximum growth rate of the coupled electroconvective and morphological instability at $\delta=10^{-4}$. $V=15$ and $V=25$ are below and above the critical voltage for the onset of the electroconvective instability at $U_m=0$. Symbols show the asymptotic solutions in equs. (\ref{eq:puremorph0}), (\ref{eq:puremorph1}) and (\ref{eq:puremorph2}) for the bulk analysis. ($b$) Maximum growth rate of the instability with  surface tension at the electrode/electrolyte interface. The bulk analysis result shows the growth rate for the purely morphological instability at large wavenumber (\ref{eq:surfbulk}) with $C|_{y=0}=0.15$.}\label{fig:fig16}
\end{center}
\end{figure}

Fig. \ref{fig:fig16}($a$) shows the growth rate of the perturbation for the coupled electroconvective and morphological instability at two different voltages $V=15$ and 25, which are below and above the critical voltage for the onset of electroconvection, respectively. At $V=15$, the full analysis agrees well with the asymptotic solutions of the bulk problem $\sigma_r\sim1$ for $k\ll1$ and $\sigma\sim k$ for $k\gg1$. The morphological instability is directly caused by the cation flux and therefore is not strongly influenced by the space charge layer. At $V=25$, the growth rate has a region where $\sigma_r\sim k^2$ due to the electroconvective instability. It reaches a local maximum at $k\simeq30$ and eventually scales as $k$ again as $k\to\infty$. This result is qualitatively different from the bulk analysis in \cite{tikekar2018electroconvection}, showing that the full analysis is necessary to correctly predict the coupled electroconvective and morphological instability. With a strong cross-flow, the local peak of the growth rate due to the electroconvective instability vanishes.

At large enough wavenumber, the surface tension of the electrode/electrolyte interface eventually becomes important and stabilizes the surface. To consider the effects of surface tension, we replace the boundary condition for the cation in equ.(\ref{eq:fbc1}) by the electrochemical potential balance at the electrolyte/electrode interface ~\cite[]{tikekar2014stability}
\begin{equation}\label{eq:surftension}
\Phi^{\theta}_{C^+}+(\Phi+\ln C^+)|_{y=h}=\Phi^{\theta}_m+\Phi_m+\gamma K,
\end{equation}
where $\Phi^{\theta}_{C^+}$ and $\Phi^{\theta}_m$ are the standard chemical potentials for the cation in the electrolyte and metal electrode, respectively, $\Phi_m$ is the electrostatic potential of the metal electrode. $\gamma=\gamma^* v^*_m/(LRT)$ is the capillary number of the depositing cation. For the lithium metal, the molar volume of the metal atom is $v^*_m=1.33\times10^{-5}\mathrm{m}^3/\mathrm{mol}$, the interfacial energy $\gamma^*=1.716\mathrm{J}/\mathrm{m}^2$, and the dimensionless interfacial energy $\gamma=9.15\times10^{-6}$. $K$ is the curvature of the electrode surface which is positive for a convex projection into the electrolyte. In the base state, equ. (\ref{eq:surftension}) reduces to the condition (\ref{eq:fbc1}) for a constant cation concentration. Here, we still assume $C^+|_{y=0}=C_s=1$.
For small perturbations, $K=-d^2h/dx^2$ and the boundary condition at the anode surface in equ. (\ref{eq:fullperturbbc1}) becomes
\begin{equation}
(c^++C^{+'}h)|_{y=0}=h\gamma k^2C^+|_{y=0}.
\end{equation}
The corresponding growth rate is shown in figure \ref{fig:fig16}($b$). The surface tension stabilizes the perturbation for $k>10^3$ and the critical wavenumber scales as $k\sim\gamma^{-1/2}$ ~\cite[]{tikekar2014stability}. Similarly for the bulk analysis, we replace the first boundary condition in equ. (\ref{eq:perturbedbc1a}) by a chemical potential balance condition for the cation $2(c|_{y=0}+h)/C|_{y=0}-\mu|_{y=0}=h\gamma k^2$. $C|_{y=0}$ is the small but non-zero cation concentration at the edge of the bulk region and here we treat it as a fitting parameter.
The growth rate in equ. (\ref{eq:puremorph2}) becomes
\begin{equation}\label{eq:surfbulk}
\sigma=(D+1)v_mk(1-C|_{y=0}\gamma k^2/2),
\end{equation}
and the mode becomes stable at $k_{cr}=\sqrt{2/(\gamma C|_{y=0})}$. In figure \ref{fig:fig16}($b$), the bulk analysis matches the full analysis for large $k$ with a fitting parameter $C|_{y=0}\sim0.15$. This result shows that the double layer and the space charge layer have significant effects on the morphological instability at high enough wavenumbers. However, it is not straightforward to directly compare the bulk and full analysis since the boundaries in the two analyses are different.   It is thus important to experimentally determine the ion concentration at the electrode/electrolyte interface to accurately predict the critical wavenumber at which the interfacial energy becomes important using the bulk analysis. To summarize, although the cross-flow cannot eliminate the morphological instability, it does affect the wavenumber of the most unstable mode by suppressing electroconvection.

%This result is consistent with the previous experimental observation of lithium dendrite growth in an electrolyte of polymer solution ~\cite[]{wei2018stabilizing}, where the dendrite changes from mossy mushroom structures ($k\sim10$) to needle-like structures ($k\sim10^3$) along with the suppression of the electroconvection.

%
%\begin{subequations}\label{eq:full}
%\begin{equation}
%\frac{\partial c^+}{\partial t}
%+Pe(\mbox{\boldmath$u$}\cdot\nabla)c^+
%=\frac{1+D}{2}\nabla\cdot(c^+\nabla\mu^+)
%\end{equation}
%\begin{equation}
%\frac{\partial c^-}{\partial t}
%+Pe(\mbox{\boldmath$u$}\cdot\nabla)c^-
%=\frac{1+D}{2D}\nabla\cdot(c^-\nabla\mu^-)
%\end{equation}
%\begin{equation}
%2U_m-\nabla p+\nabla^2\mbox{\boldmath$u$}
%+\nabla^2\phi\nabla\phi=0, \quad \nabla\cdot\mbox{\boldmath$u$}=0,
%\end{equation}
%\begin{equation}
%-2\delta^2\nabla^2\phi
%=c^+-c^-,
%\end{equation}
%\begin{equation}
%\phi=(\mu^+-\mu^--\ln c^++\ln c^-)/2,
%\end{equation}
%\end{subequations}

\section{Conclusion}\label{sec:conclusion}

We studied the effects of cross-flow on the electroconvective and morphological instabilities in an electrolyte near an ion-selective surface using two methods. In the bulk analysis, we use the electro-osmotic slip velocities as the boundary conditions for the bulk region and derive the asymptotic solutions for small and large wavenumbers. In the full analysis, we numerically calculate the base state as well as the perturbed solution using the ultraspherical spectral method. In both studies, the general effect of the cross-flow is to attenuate the influences of the electroconvection by suppressing the vortices at small wavenumbers near the ion-selective surface. For the purely electroconvective instability, the imposed flow generates a stable central mode below a critical wavenumber, and decreases the growth rate of the wall mode at large wavenumbers. The transition wavenumber between the two modes scales as $k\sim U_m^{1/2}$. In the full analysis, since the high wavenumber mode is stabilized by the diffusion of ions inside the double layer and space charge layer, a complete suppression of electroconvection is achievable by imposing a critical flow rate whose amplitude is larger for higher voltages and thinner double layers.

In spite of the imposed flow, a metal electrode adjacent to a Newtonian electrolyte is always subject to the morphological instability. Its growth rate scales as $\sigma\sim1$ for $k\ll1$ and $\sigma\sim k$ for $k\gg1$. The onset of the electroconvective instability greatly increases the growth rate of the coupled instability to $\sigma\sim k^2$ and generates a local maximum at a moderate wavenumber ($k\sim10$).
The imposed flow increases the growth rate at small wavenumber by increasing the ion concentration gradient near the electrolyte/electrode interface. At large wavenumber, the flow reduces the growth rate mainly by suppressing the electroconvective instability. This result shows that the imposed flow can change the morphology of the electrodeposition from mossy-like to needle-like structures.

The comparison between the analytical treatment of the bulk region and the full stability analysis shows that the analytical result  must be used with caution. For the purely electroconvective instability, the bulk analysis significantly underestimates the critical voltage and fails to predict the transition from unstable to stable modes with increasing wavenumber. When the applied voltage is well above the critical voltage for the onset of the electroconvective instability, the transition wavenumber from the full calculation scales as the inverse of the thickness of the space charge layer $k\sim\delta^{-2/3}$, a result that might be anticipated by applying a simple cut off to the bulk analysis. When the applied voltage is close to the critical voltage, however, no such relation was found, suggesting a more complicated interaction between the perturbation and the space charge layer which can only be captured by the full analysis. For the electroconvective instability without the flow, the perturbed velocity transports more ions from regions of low concentration to regions of high concentration than predicted by the bulk analysis. Once the flow is strong enough to overcome the stabilizing effects caused by ion diffusion and migration, it generates a positive feedback and causes the electroconvective instability. By studying the transition from unstable to stable modes at various double layer thicknesses and applied voltages, our result shows that the base state becomes stable when the maximum normal velocity $v_{max}<0.22k^2V$ for an ion concentration with a unitary peak. The stabilizing effect of the imposed flow is not caused by reducing the perturbation velocity. In fact, the magnitude of the perturbation velocity increases with increasing magnitude of the imposed flow. Instead, the shear flow deforms the perturbed ion concentration field and generates a shielding effect which suppresses the ion flux from low to high concentration regions. The bulk analysis qualitatively captures these features but the results are quantitatively different from those obtained with the full analysis. For the purely morphological instability, the bulk analysis predicts accurate result for both small and large wavernumbers. However, when the surface tension becomes important, the ion concentration distribution near the ion-selective surface, which must be determined from the full analysis or experimental measurements, significantly affects the prediction of the critical wavenumber.

\begin{acknowledgments}
This research is supported by Department of Energy Basic Energy Science Grant No. DE-SC0016082.
\end{acknowledgments}

\bibliography{ref}

\end{document}